%% file: main.tex
\documentclass[sort&compress,preprint]{elsarticle}
\usepackage{geometry}
\usepackage{setspace}
\usepackage[absolute,overlay]{textpos}
\usepackage{amsmath,graphicx} 
\usepackage{siunitx}
\usepackage{comment}
\usepackage{braket}
\usepackage{amssymb}
\usepackage{gensymb}
\usepackage{mathtools}
\usepackage[thinc]{esdiff}
\usepackage[capitalise]{cleveref}
\usepackage{xfrac}
\usepackage[dvipsnames]{xcolor}
\usepackage[normalem]{ulem}
\usepackage{multirow}
\usepackage{leftidx}
\usepackage{mathrsfs}

\usepackage{float}
\usepackage{siunitx}

\journal{Journal of Nuclear Materials}

\usepackage{soul}
\soulregister\emph7 
\soulregister\cite7
\soulregister\ref7

\usepackage{subfloat}
\usepackage{caption}
\usepackage{subcaption}
\usepackage{graphicx}
\graphicspath{{Figures/}}

\begin{document}

\begin{frontmatter}
\onecolumn
\title{Influence of temperature, initial grain-boundary bubble density and grain structure on fission gas behaviour in UO$_2$: a 3D hybrid multiscale study}
\author[inst1,inst5]{Sourav Chatterjee}

\affiliation[inst1]{organization={University of Florida},
            city={Gainesville},
            postcode={32611}, 
            state={FL},
            country={USA}}
\author[inst1,inst2]{Md Ali Muntaha}
\author[inst3]{Sophie Blondel}
\author[inst4]{David Andersson}
\author[inst3]{Brian D. Wirth}
\author[inst1]{Michael R. Tonks}

\affiliation[inst2]{organization={Idaho National Laboratory},
            Department and Organization = {Fuel Development Performance and Qualification},
            city={Idaho Falls},
            state={ID},
            country={USA}}

\affiliation[inst3]{organization={University of Tennessee},
            city={Knoxville},
            state={TN},
            country={USA}}
            
\affiliation[inst4]{organization={Los Alamos National Laboratory},
            city={Los Alamos},
            state={NM},
            country={USA}}

\affiliation[inst5]{organization={Lawrence Livermore National Laboratory},
            Department and Organization = {Materials Science Divison},
            city={Livermore},
            postcode={94550}, 
            state={CA},
            country={USA}}

\begin{abstract}Fission gas swelling and release in UO$_2$ are governed by the coupled evolution of intragranular clusters and bubbles, migrating grain boundaries (GBs), triple junctions (TJs), and their eventual connection to a free surface (FS). We extend a hybrid multiscale framework that couples cluster dynamics (Xolotl) with a phase‑field model (MARMOT) to large 3D polycrystals with heterogeneous GB and surface diffusion and evolving GB networks. We simulate 10‑ and 100‑grain UO$_2$ microstructures at 1200 and 1600 K, with and without a FS, to interrogate bubble growth, coalescence, GB/TJ coverage, gas arrival at interfaces, and fission gas release (FGR). At 1200 K, both GB mobility and gas transport are low, yielding negligible bubble and GB evolution. At 1600 K, intergranular bubbles rapidly become lenticular and coalesce into networks while unpinned GBs migrate; fewer initial bubbles reduce coalescence but enhance GB migration due to less pinning and produce spikes in interfacial gas arrival rate due to GB sweeping. Bubble density versus mean projected area agrees with White’s (2004) coalescence trend and remains on the left side of the analytical curve, in contrast to several prior simulations, likely due to the inclusion of GB migration. In domains with a FS, early release is rapid and bubbles near the FS collapse to form a denuded zone, suppressing local network connectivity; GB coverage rises and approaches but does not exceed 50\%. TJ coverage remains low without preferential nucleation at TJs. To our knowledge, these are the first large-scale 3D mesoscale simulations of intergranular fission gas behavior that provide mechanistic insight and quantitative metrics to inform engineering‑scale FGR models.
\end{abstract}

\begin{keyword}
Phase field modeling \sep Multiscale modeling \sep MOOSE-Xolotl coupling \sep Fission gas release \sep Microstructure evolution
\end{keyword}

\end{frontmatter}

\input{Introduction}

\input{Model}

\input{Results}

\newpage
\input{Appendix_A}

\section*{Credit Authorship Contribution Statement}
\textbf{Sourav Chatterjee}: Investigation, Methodology, Analysis, Software, Writing – original draft. \textbf{Md Ali Muntaha}: Investigation, Methodology, Analysis, Software,
Writing – original draft. \textbf{Sophie Blondel}:  Investigation, Methodology, Analysis, Writing – review and editing.  \textbf{David Andersson}:  Conceptualization, Funding acquisition, Project administration, Supervision, Writing – review and editing. \textbf{Brian Wirth}: Conceptualization, Project administration, Supervision, Writing – review and editing. \textbf{Michael R. Tonks}: Conceptualization, Funding acquisition, Project administration, Supervision, Writing – review and editing.

\section*{Declaration of Competing Interest}
The authors declare that they have no known competing financial interests or personal relationships that could have appeared to influence the work reported in this paper.

\section*{Acknowledgements}
This material is based upon work supported by the U. S. Department of Energy, Office of Nuclear Energy and Office of Science, Office of Advanced Scientific Computing Research through the Scientific Discovery through Advanced Computing (SciDAC) project on Simulation of Fission Gas through the grant DOE DE-SC0018359 at the University of Tennessee.

This research made use of the resources of the High-Performance Computing (HPC) Center at Idaho National Laboratory, which is supported by the Office of Nuclear Energy of the U.S. Department of Energy and the Nuclear Science User Facilities under Contract No. DE-AC07-05ID14517. 

Part of this work was performed under the auspices of the US Department of Energy by Lawrence Livermore National Laboratory (LLNL) under Contract DE-AC52-07NA27344 (SC). The authors would like to thank LLNL for providing the HPC resources to perform the free-surface simulations.

Los Alamos National Laboratory, an affirmative action/equal opportunity employer, is operated by Triad National Security, LLC, for the National Nuclear Security Administration of the U.S. Department of Energy under contract number 89233218CNA000001.

\section*{Data Availability}
The data and MOOSE input files that support the findings of this study are available from the authors upon reasonable request.

\newpage
\bibliographystyle{elsarticle-num.bst}
\bibliography{paper5.bib}

\end{document}

%% file: Introduction.tex
\section{Introduction}
\label{sec:introduction}
Uranium dioxide (UO$_2$) pellets stacked inside zirconium cladding tubes are commonly used as reactor fuel in commercial Light Water Reactors (LWRs)~\cite{Olander, rest2019fission, tonks}. As fissioning occurs within the fuel, nearly 30\% of the fission products are noble gases - primarily xenon (Xe) and krypton (Kr), which are essentially insoluble in UO$_2$~\cite{Olander, rest2019fission, tonks}. Consequently, these fission gas atoms precipitate as intra- and intergranular gas bubbles~\cite{Olander, rest2019fission, tonks}. These gas bubbles reduce the thermal conductivity of the fuel and cause swelling~\cite{Olander, tonks}. Fuel swelling decreases the gap between the fuel and the cladding and eventually results in a fuel-cladding interaction, potentially leading to cladding failure~\cite{Olander, rest2019fission, tonks}. Over time, intergranular gas bubbles grow and coalesce to form interconnected pathways that eventually connect to a free surface (FS), allowing fission gases to be released from the fuel. Fission gas release (FGR) further degrades fuel performance by reducing the thermal conductivity across the fuel-cladding gap and increasing the plenum pressure~\cite{Olander, rest2019fission, tonks}. 

It has been proposed that FGR in UO$_2$ occurs in three intrinsically coupled stages~\cite{tonks}. The first stage involves the production of fission gas atoms, their diffusion toward UO$_2$ grain boundaries (GBs), their trapping by intragranular bubbles and irradiation-induced defects, and their re-solution into the UO$_2$ matrix~\cite{Olander, rest2019fission, tonks}. The second stage involves the formation, growth, and interconnection of lenticular-shaped gas bubbles on the GBs to form interconnected networks that can reach triple junctions (TJs)~\cite{tonks}. The third stage is the formation and interconnection of fission gas tunnels along TJs until they reach a FS, allowing the gas contained within the bubbles to escape. All three stages happen simultaneously within the fuel. Experimental evidence suggests that the intergranular bubbles formed during Stages Two and Three are considerably larger (with radii on the order of microns) than the intragranular bubbles formed during the first stage (with radii on the order of nanometers), thus they retain a significant amount of fission gas atoms and contribute substantially to fuel swelling~\cite{Olander, tonks}. Owing to the strong coupling between the three stages, predicting FGR and bubble-induced fuel swelling requires a multiscale approach capable of resolving the production, diffusion, trapping, clustering, and re-solution of fission gas atoms within UO$_2$ grains, while simultaneously handling the evolution of intergranular bubbles and migration of UO$_2$ GBs.  

Analytical physics-based models~\cite{booth, Beere, cornell_1969,  Koo, Massih, speight_1969, Tucker_75, Tucker_78, Tucker_79, Tucker_80, Forsberg, Kogai_97, Turnbull, White_83, Uffelen, White_2004}, starting from the Booth model \cite{booth}, have been proposed to predict both FGR and fuel swelling in UO$_2$. Pastore~\textit{et al.}~\cite{Pastore_2013} proposed a mechanistic model that predicts both FG swelling and release. However, these models make simplifying assumptions about the interaction of intra- and intergranular bubbles and GBs and how they eventually result in FGR. It is impossible to know the accuracy of these assumptions using reactor fuel data since it is not currently possible to watch fission gas bubble evolution within the fuel in situ during reactor operation. 

To test these assumptions using simulations, several researchers~\cite{hu2009phase, Millet_2012_a, Millet_2012_b, Millet_2012_c, Larry_2019, PRUDIL2022153777, lan2024three} have utilized the phase-field approach to model the behavior of fission gas bubbles in UO$_2$. However, it is computationally intractable to apply the phase-field approach to simultaneously model both the trapping of fission gas atoms by nanoscale intragranular gas bubbles and the evolution of microscale intergranular bubbles. Therefore, this approach can be used to model intragranular bubbles \cite{li2013phase,tonks2013comparison} or intergranular bubbles~\cite{Millet_2012_a, Larry_2019, PRUDIL2022153777, lan2024three} but not both. 

Recently, Kim \textit{et al.}~\cite{DongUk-Kim} addressed this limitation by coupling a phase-field model with a spatially resolved cluster dynamics model. The phase-field model handles the evolution of both UO$_2$ GBs and intergranular bubbles at the microscale, while the cluster dynamics model accounts for the production, diffusion, trapping, clustering and re-solution of fission gas atoms within the UO$_2$ grains. This hybrid model assumed homogeneous gas diffusivity~\cite{DongUk-Kim}, an assumption that is not accurate for polycrystalline UO$_2$ microstructures. Subsequently, Muntaha~\textit{et al.}~\cite{muntaha2023} modified the hybrid model to include fast GB and surface diffusion of fission gas atoms and vacancies, and predicted the amount of fission gas released using 2D simulations. 

Another limitation of current PF studies is their use of either 2D simulations or small 3D simulations to investigate fission gas bubble behavior. PF models of intragranular bubble evolution have shown that 3D simulations are required to accurately represent experimental behavior \cite{li2013phase,tonks2013comparison}. 2D simulations of GB and intergranular bubble evolution can never accurately represent FGR because TJs are points rather than lines and therefore cannot result in the formation of TJ tunnels during Stage Three of FGR~\cite{muntaha2023}. 3D bicrystal simulations were used to understand the interconnection of intergranular bubbles \cite{Millet_2012_a,lan2024three}, but could not represent TJs since they included only two grains. Aagesen~\textit{et al.}~\cite{Larry_2019} investigated the behavior of TJ bubbles in a 3D hexagonal grain structure with four grains; however, the grain structure did not evolve and did not include a FS for FGR. Prudil~\textit{et al.}~\cite{PRUDIL2022153777} used the included-phase model to represent the third dimension indirectly and modeled the bubble evolution in large static polycrystals without a FS for release. In our past work~\cite{muntaha2023}, we modeled the evolution of FG bubble in a closed 3D five-grain polycrystals domain, but we only simulated 75 min of evolution since our focus was on determining a reasonable vacancy surface diffusivity rather than analyzing the bubble behavior and FGR. 3D simulations of around five days of evolution in polycrystals with 4, 8, 16 and 32 columnar grains were carried out by Lan ~\textit{et al.}~\cite{lan2024three}, but their TJs were small and always parallel to the z-axis and they did not consider a FS for FGR.

The purpose of this work is to use the hybrid fission gas model~\cite{DongUk-Kim,muntaha2023} to investigate the interactions between GBs, intergranular and TJ bubbles, and a FS in large 3D polycrystal simulations. Specifically, we first consider a $10$-grain UO$_2$ polycrystal to carry out a focused analysis of the growth and interconnection of intergranular bubbles, their interaction with GBs, and the formation of TJ bubbles. Since GB mobilities and diffusivities vary with temperature, we perform simulations at  $1200$ and $1600$ K. We validate our model predictions by comparing the evolution behavior of intergranular bubbles with White's experimental result \cite{White_2004}. We then simulate a $100$-grain polycrystal to get a broader view of bubble and GB evolution. These simulations are the first large-scale 3D polycrystal fission gas bubble simulations that resolve both intragranular and intergranular bubbles.

We begin by summarizing the hybrid model in \cref{sec:formulation}. We then present our 10-grain and 100-grain simulation results in \cref{sec:results}. We discuss the results in \cref{sec:discussion}, and conclude in \cref{sec:conclusions}.

%% file: Model.tex
\section{Hybrid multiscale model description}
\label{sec:formulation}
In this section, we briefly summarize our hybrid multiscale model \cite{DongUk-Kim,muntaha2023}. By combining a cluster dynamics model with a PF model based on the work by Aagesen \emph{et al.}~\cite{Larry_2019}, the hybrid model predicts the evolution of both intergranular bubbles and UO$_2$ GBs at the microscale, while accounting for the production, diffusion, trapping and re-solution of fission gas atoms within UO$_2$ grains. For a more detailed description of these models, refer to Refs.~\cite{Larry_2019}, \cite{DongUk-Kim}, \cite{muntaha2023}.

\subsection{Cluster dynamics model}
\label{CD_model}
The cluster dynamics model includes production, diffusion, trapping, and re-solution of fission gas atoms, specifically Xe~\cite{DongUk-Kim}. In this model, the concentration of an intragranular Xe cluster containing $n$ atoms evolves according to the following mass conservation equation \cite{DongUk-Kim}:
\begin{equation}
    \frac{\partial C_n}{\partial t} = D_n \nabla^2 C_n + \dot{F}y_n - Q(C_n),
      \label{DC}
\end{equation}
where $C_n$ is the cluster concentration,  $D_{n}$ is the corresponding diffusion coefficient, $\dot{F}$ is the fission rate density, and $y_{n}$ is the fission yield of the cluster.  In this model, we assume that only single Xe atoms diffuse, such that $D_{n}=0$ for $n\geq2$ \cite{DongUk-Kim}. Further, because fission reactions yield only single Xe atoms, $y_{n}=0$ for $n\geq2$~\cite{DongUk-Kim}. For clusters of size $n\geq 2$, $Q(C_n)$ describes the trapping, emission, and re-solution of Xe gas atoms within UO$_2$ grains and is expressed as follows~\cite{DongUk-Kim}:
\begin{equation}
Q(C_{n}) = k_{n}C_{n}C_{1} - k_{(n-1)}C_{(n-1)}C_{1} + k_{n}^{emit}C_{n} -k_{(n+1)}^{emit}C_{(n+1)} + k_{n}^{reso}C_{n} - k_{(n+1)}^{reso}C_{(n+1)},
\end{equation}
with 
\begin{align}
\begin{split}
k_{n} &= 4\pi D_{1}(r_{1} + r_{n})\\
k_{(n+1)}^{emit} &= k_{n}/v_{a}\,\mathrm{exp}(-E_{b}/k_{B}T)\\
k_{n}^{reso} &=\left[a_{1}\mathrm{exp}(-b_{1}r_{n}) + \frac{y(0) - a_{1}}{1+c r_{n}^2}\mathrm{exp}\left(-b_{2}r^{2}_{n}\right) \right]10^{4}\dot{F}.
\end{split}
\end{align}
Here, $D_{1}$ and $r_{1}$ are the diffusion coefficient and reaction radius of a single Xe atom, respectively. $r_{n}$ is the reaction radius of a cluster with $n$ Xe atoms, $v_{a}$ is the atomic volume of UO$_2$, $E_{b}$ is the binding energy, $k_{B}$ is the Boltzmann constant, and $T$ is temperature. 

The values adopted for solving Eq.~\eqref{DC} are summarized in Table~\ref{table1}. It is important to note that the cluster dynamic model passes all clusters to the PF model at UO$_2$ GBs and Xe-gas bubble surfaces, such that the concentration of Xe clusters is zero at these boundaries~\cite{DongUk-Kim}. In addition, in simulations with a FS, the concentration of all Xe clusters at the surface is released from the system, such that their concentration at the surface is also zero~\cite{muntaha2023}.

Furthermore, we have used two re-solution models. The simulations with a FS are performed using the heterogeneous re-solution model of Setyawan \emph{et al.}~\cite{setyawan2018atomistic}, where a Xe bubble ejects an individual Xe atom at a time. However, in the periodic cases, we employ the Turnbull re-solution model, as described in Eqs.~(2) and (16) of the work by Pastore \emph{et al.}~\cite{pastore2023}, in which the Xe bubble is fully re-solved in the lattice. The latter leads to a higher concentration of mobile Xe in the grains and a larger amount of Xe moving to the GB, leading to much larger intragranular bubble growth than in the former.

\subsection{Phase-field model}
\label{PF_Model}
\subsubsection{Governing equations}
The PF model handles the evolution of both UO$_2$ GBs and intergranular bubbles. Following Refs.~\cite{Larry_2019}, \cite{DongUk-Kim}, \cite{muntaha2023}, the UO$_2$ grains are denoted by a set of order parameters $\boldsymbol{\eta}_{u} =$ $\lbrace \eta_{u1}, \eta_{u2}, ..., \eta_{uN}\rbrace$, where $N$ is the number of grains. The order parameter $\eta_{ui}$ assumes a value of $1$ within the grain it represents and is $0$ in all other grains and in the intergranular bubble regions. It smoothly transitions from $0$ to $1$ within the GBs, such that $0<\eta_{ui}<1$. Similarly, the intergranular bubbles are denoted by an additional order parameter $\eta_{b0}$, which is $1$ within the bubbles and is $0$ elsewhere. To decouple the bulk and interfacial properties, we derive the evolution equations of the order parameters starting from the following grand-potential functional \cite{Plapp_2011,Larry_2019,DongUk-Kim,muntaha2023}:
\begin{equation}
    \Omega =\int_{V} \left[ mf_{0}\left(\boldsymbol{\eta}_{u}, \eta_{b0}\right) + \frac{\kappa}{2} \left(\sum_{i=1}^{N} |\nabla \eta_{ui}|^2  + |\nabla\eta_{b0}|^2\right)+ h_{m}\omega_{m} + h_{b}\omega_{b} \right]dv,
  \label{eq:omega}
\end{equation}
with $\omega_{m}$ and $\omega_{b}$ denoting the bulk grand-potential densities of the matrix and bubble phases, respectively. The parameters $\kappa$ and $m$ denote the gradient energy coefficient and barrier height, respectively. These parameters are related to the GB energy $(\sigma)$ and an assumed interface width ($l_{w}$) according to $\kappa = (3/4) \sigma  l_{w}$ and $m=6\sigma /l_{w}$~\cite{Moelans_2011}.  The function $f_{0}\left(\boldsymbol{\eta}_{u}, \eta_{b0}\right)$ is formulated as by Moelans~\cite{Moelans_2011}:
\begin{equation}
    f_{0}\left(\boldsymbol{\eta}_{u}, \eta_{b0} \right) = \sum_{i=1}^{N}\left(\frac{\eta_{ui}^{4}}{4} - \frac{\eta_{ui}^{2}}{2}\right)  + \left(\frac{\eta_{b0}^{4}}{4} - \frac{\eta_{b0}^{2}}{2}\right)+ \gamma\left(\eta_{b0}^{2}\sum_{i=1}^{N}\eta_{u i}^{2} + \sum_{i=1}^{N}\sum_{j=i+1}^{N}\eta_{ui}^2\eta_{uj}^2\right) + \frac{1}{4},
\end{equation}
where $\gamma$ is a model parameter that controls the width of the interface $l_w$. The interpolation (or switching) functions corresponding to the matrix ($h_m$) and bubble ($h_b$) phases are expressed as follows \cite{Larry_2019, Moelans_2011}: 
\begin{align}
    h_m(\boldsymbol{\eta}, \eta_{b0}) =\frac{\sum_{i=1}^{N}\eta_{ui}^2}{{\sum_{i=1}^{N}\eta_{ui}^2} + \eta_{b0}^2} \quad \text{and} \quad h_{b}(\boldsymbol{\eta}, \eta_{b0})  = \frac{\eta_{b0}^2}{{\sum_{i=1}^{N}\eta_{ui}^2} + \eta_{b0}^2}
  \end{align}
As presented by Plapp~\cite{Plapp_2011}, the bulk grand potential densities corresponding to the matrix $(\omega_{m})$ and bubble $(\omega_{b})$ phases are obtained by Legendre transform of the free energy densities, which yields:
\begin{align}
    \omega_\theta\left(\mu_{g},\mu_{v}\right) = f_{\theta} - \mu_g \left(c_g^{\theta}/v_{a}\right) - \mu_v \left(c_v^{\theta}/v_a\right), \quad \theta = \lbrace m, b\rbrace,
    \label{Eqn5}
\end{align}
where $v_{a}$ is the atomic volume,  $\mu_{g}$ and $\mu_{v}$ are the chemical (or diffusion) potentials of Xe and uranium vacancies (Va), respectively, and $f_{\theta=m,b}$ are the bulk free energy densities of the matrix and bubble phases. The terms $c_g^{\theta}(\mu_{g})$ and $c_v^{\theta}(\mu_{v})$ denote the phase atomic fractions of Xe and uranium Va within the matrix and bubble phases, respectively, and are functions of chemical potential \cite{Plapp_2011}. Minimization of Eq.~\eqref{eq:omega} with respect to the chemical potentials yields the following equation: 
\begin{align}
    c_{k} &=v_{a}\left(h_{m}\frac{\partial \omega_{m}}{\partial \mu_{k}} +  h_{b}\frac{\partial \omega_{b}}{\partial \mu_{k}}\right) = h_{m}c_{k}^{m}(\mu_{k}) +  h_{b}c_{k}^{b}(\mu_{k}),\quad k=\{g, v\},
    \label{Eqn5a} \quad k=\{g, v\},
\end{align}
with $c_{k=g,v}$ denoting the overall atomic fractions of Xe and Va. Assuming the overall gas and uranium vacancies are conserved quantities, their evolution is governed by the following mass conservation equations:
\begin{align}
    \frac{1}{v_{a}}\frac{\partial c_{k}}{\partial t}= \nabla(\mathcal{M}_k \nabla\mu_k) +  s_{k} \quad k=\{g, v\},
    \label{Eqn5c}
\end{align}
where $\mathcal{M}_{k=g,v}$ are the atomic mobilities of Xe and Va and $s_{k=g,v}$ are the source terms accounting for the production of gases and uranium vacancies. It is important to note that the gas source term $s_{g}$ is a spatially varying quantity that is passed from the cluster dynamics model to the PF model \cite{DongUk-Kim}. Unlike the gas production term $(\dot{F}y_{n})$ in Eq.~\eqref{DC}, it is active only near the UO$_2$ GBs and bubble surfaces \cite{DongUk-Kim}. This is because the source term in Eq.~\eqref{DC} comprises all the fission gas atoms generated during irradiation within UO$_2$ grains, while the source term in Eq.~\eqref{Eqn5c} comprises the mobile gas atoms that have either diffused to GBs or intergranular bubble surfaces or are swept by GB migration~\cite{DongUk-Kim}. Since Xolotl does not model uranium vacancies, the vacancy production rate $s_{v}$ in Eq.~\eqref{Eqn5c} is assumed to be four times the gas production rate $s_{g}$, i.e., $s_{v}=4s_{g}$~\cite{DongUk-Kim,muntaha2023}, since Xe atoms in UO$_2$ diffuse with four vacancies under irradiation at the simulated temperatures of interest \cite{Matthews_2020}. 

By differentiating Eqs.~\eqref{Eqn5a} with respect to time and using Eqs.~\eqref{Eqn5c}, the mass conservation equations can be alternatively written as chemical potential evolution equations, as demonstrated in previous studies~\cite{Plapp_2011, Larry_2019}. This yields the following chemical potential evolution equations :
\begin{align}
    \chi_{k}\frac{\partial \mu_{k}}{\partial t}= 
    \nabla(\mathcal{M}_k\nabla\mu_k) + 
    s_k -
    \frac{\partial h_{m}}{\partial t}\left(\frac{c_{k}^{m}- c_{k}^{b}}{v_{a}}\right),\quad k=\{g, v\},
     \label{Eq:chem_g_evolution}
\end{align}
with
\begin{align}
    \frac{\partial h_{m}}{\partial t} &= \sum_{i=1}^{N}\frac{\partial h_{m}}{\partial \eta_{ui}}\frac{\partial \eta_{ui}}{\partial t} + \frac{\partial h_{m}}{\partial \eta_{b0}}\frac{\partial \eta_{b0}}{\partial t},\\
    \chi_{k}(\boldsymbol{\eta}, \eta_{b0}) &= \left(\frac{h_{m}}{v_{a}^2\mathcal{A}_{k}^{m}} + \frac{h_{b}}{v_{a}^2\mathcal{A}_{k}^b} \right), \quad k=\{g, v\}.
\end{align}
Minimization of Eq.~\eqref{eq:omega} with respect to the non-conserved order parameters yields the following evolution equations for the order parameters corresponding to the grains and bubble phases:
\begin{align}
    \frac{ \partial{\eta_{ui}} }{\partial{t}} &= -L\left[m\left(\eta_{ui}^{3} - \eta_{ui}\right) + 2m\gamma\eta_{ui}\left(\eta_{bo}^2 + \sum_{i=1}^{N}\eta_{ui}^2\right) + \kappa \Delta \eta_{ui} + 
    \frac{\partial h_{m}}{\partial \eta_{ui}}\left(\omega_{m} -\omega_{b}\right) \right],\quad i=\lbrace1\hdots N \rbrace 
      \label{Eq:AC1_evolution}
      \\
    \frac{ \partial{\eta_{b0}} }{\partial{t}} &= -L\left[m\left(\eta_{b0}^{3} - \eta_{b0}\right) + 2m\gamma\eta_{b0}\sum_{i=1}^{N}\eta_{ui}^2 + \kappa \Delta \eta_{b0} + \frac{\partial h_{m}}{\partial \eta_{b0}}\left(\omega_{m} - \omega_{b}\right) \right], 
     \label{Eq:AC2_evolution}
\end{align}
where $\Delta$ denotes the Laplacian operator. $L$ is the mobility of the order parameters, which is defined as follows \cite{greenquist2020development}: 
\begin{align}
    L\left(\eta_{b0}\right) = L_{m} \left[1- H\left(\eta_{b0}\right)\right]  + L_{b} H (\eta_{b0}),
    \label{Eqn11}
\end{align}
where $L_{m}$ denotes the mobility of the order parameters corresponding to the matrix phase, $L_{b}$ denotes the mobility of the order parameter corresponding to the bubble phase, and $H$ is an interpolation function. The function $H$ is formulated as follows:
\begin{align}
    H\left(\eta_{b0}\right) = 
    \begin{cases}
        0 \quad  &\text{if}\quad \eta_{b0} \leq 0\\
        6\eta_{b0}^3 - 15\eta_{b0}^2 + 10\eta_{b0}^2 \quad &\text{if}\quad  0<\eta_{b0}<1\\
        1 \quad  & \text{if}\quad \eta_{b0} \geq 1
    \end{cases}
\end{align}
The GB mobility is related to the mobility of the order parameters according to~\cite{muntaha2023}:
\begin{align}
    L_{m} = \frac{4}{3}\left(\frac{M_{0} e^{-Q/k_bT}}{l_{w}}\right),
\end{align}
where $M_{0}$ is the GB mobility prefactor and $Q$ is the GB mobility activation energy. $L_{b}$ in Eq.~\eqref{Eqn11} is assumed to be $10$ times greater than $L_{m}$ at any given temperature so that the migration of the intergranular bubble interface is diffusion-controlled rather than interface-controlled \cite{muntaha2023}.

\subsubsection{Heterogeneous GB and surface diffusion}
Compared to previous works \cite{Larry_2019, DongUk-Kim,lan2024three}, a key distinguishing feature of our model is that we include heterogeneous gas and vacancy diffusion within the UO$_2$ microstructure. Following Muntaha \emph{et al.}~\cite{muntaha2023}, this is incorporated in Eq.~\eqref{Eq:chem_g_evolution}, where atomic mobilities $\mathcal{M}_{k}$ are related to diffusivities \cite{Plapp_2011} according to $\mathcal{M}_{k} = D_{k}\chi_{k}\ \mathrm{for}\, k=\{g,v\}$. Since GB and surface diffusivities are several orders of magnitude faster than bulk in polycrystalline UO$_2$ \cite{muntaha2023}, the chemical diffusivities $D_{k}$ are formulated as functions of the order parameters \cite{muntaha2023}:
\begin{align}
D_{k}\left(\boldsymbol{\eta}_{u}, \eta_{b0}\right) = D_{k}^{m}h_{m} + D_{k}^{b}h_{b} + h_{gb}  \left(D_{k}^{gb} - D_{k}^m\right) + h_{s}\left(D_{k}^s - D_{k}^m\right) ,\quad k=\{g, v\},
\end{align}
where $D_{k=g,v}^{m}$ and $D_{k=g,v}^{b}$ are the diffusivities of Xe and Va within the bulk UO$_2$ matrix and intergranular bubbles, respectively. Likewise, $D_{k=g,v}^{gb}$ and $D_{k=g,v}^s$ are the diffusivities of Xe and Va along the UO$_2$ GBs and bubble surfaces, respectively. The interpolation functions representing the GBs $(h_{gb})$ and surfaces $(h_{s})$
are formulated as follows \cite{muntaha2023}:  
\begin{align}
h_{gb}\left(\boldsymbol{\eta}_{u}\right) &= 16\sum_{i=1}^{N}\sum_{j=i+1}^{N}\eta_{ui}^2\eta_{ui}^2\\
h_{s}\left(\eta_{b0}\right) &= 16\eta_{b0}^2 (1-\eta_{b0})^2. 
\end{align}
Generally, the assumed interface width $l_{w}$ in a PF model is chosen to be significantly higher than the actual width of a GB ($w_{gb}$) or a bubble surface $(w_{s})$ to reduce computational costs. Consequently, without any correction, our model would overpredict the transport of gas and vacancies along the GBs and bubble surfaces. To limit this effect, we modify the diffusivities as follows \cite{hou2019study,muntaha2023}:
\begin{align}
D_{k}\left(\boldsymbol{\eta}_{u}, \eta_{b0}\right) = D_{k}^{m}h_{m} + D_{k}^{b}h_{b} + \frac{w_{gb}}{l_{w}} h_{gb}\left(D_{k}^{gb} - D_{k}^m\right) + \frac{w_{s}}{l_{w}}h_{s}\left(D_{k}^s - D_{k}^m\right) ,\quad k=\{g, v\}.
\end{align}
In our previous work \cite{muntaha2023}, we have pointed out that there is large variability in the surface and GB diffusivity values reported in the literature. Based on the amount of fission gas released and microstructure evolution in 2D and 3D, we have suggested reasonable values for surface and grain boundary diffusivities \cite{muntaha2023}. In this work, we use these suggested values to develop a more quantitative understanding of Xe gas bubble and GB evolution using 3D simulations.

The grand-potential densities $\omega_{\theta=m,b}$ required in Eqs.~\eqref{Eq:AC1_evolution}-\eqref{Eq:AC2_evolution} are given in Appendix~\ref{AppendixA1}. The phase atomic fractions $c_{g}^{\theta}$ and $c_{v}^{\theta}$ required in Eqs.~\eqref{Eq:chem_g_evolution} are also provided in Appendix~\ref{AppendixA1}. Table~\ref{table1} provides the remaining model parameters required to solve Eqs.~\eqref{Eq:chem_g_evolution}, \eqref{Eq:AC1_evolution}, and \eqref{Eq:AC2_evolution}.

\subsection{Numerical implementation and code coupling}
\begin{table}[btp]
\caption{Model parameters used for the cluster dynamics model and PF model.}
\begin{center}
\begin{tabular}{cccc}
\hline
Parameter  & Value(s)  &  Units & Reference(s)\\ 
\hline
$T$              & $1200$ \& $1600$    & K & -----\\
$k_{B}$        &   $1.38 \times 10^{-23}$   & J/K         & ----- \\
$v_{a}$        &   $0.0409$                      & nm$^3$         & \cite{Idiri_2004} \\
$\dot{F}$ &   $3.27 \times 10^{19} $   & \text{atoms}/(m$^3$s)& \cite{Olander} \\ [1ex]
\multicolumn{4}{l}{\textit{Cluster dynamics model parameters}}\\[1ex]
$r_{1}$ &0.3&nm&  \cite{DongUk-Kim}  \\
$y_{1}$ &0.2156&----& \cite{International_Atomic_Energy_2017}  \\
$y(0)$ & $9.1816\times 10^{-4}$ & 1/s & \cite{DongUk-Kim} \\
$a_{1}$ & $0.949\times 10^{-4}$ & 1/nm &  \cite{DongUk-Kim} \\
$b_{1}$ & $0.703$ & 1/nm & \cite{DongUk-Kim} \\
$b_{2}$ & $0.0371$ & 1/nm & \cite{DongUk-Kim} \\
$c$ & $7.9821$ & 1/nm$^2$ &  \cite{DongUk-Kim}\\[1ex]
$D_{1}$ & $D_g^m$ & m$^2$/s & \cite{Turnbull, pizzocri2018model}\\[1ex]
\multicolumn{4}{l}{\textit{PF model parameters}}\\[1ex]
$\sigma$      &   $1.5$                           & J/m$^2$          & \cite{Tonks-PC-Jake_2021} \\
$M_{0}$       &    $2.14\times10^{-7}$    & m$^4$/{Js}    & \cite{Tonks-PC-Jake_2021} \\
$Q$              &   $290$                          &  kJ/{mol}         & \cite{Tonks-PC-Jake_2021} \\
$l_{w}$         &  $480$ \& $960$    	  & nm                   & -----\\
$w_{gb}$      &   $0.5$  		            & nm                  & \cite{Yao_2017}  \\
$w_{s}$        &   $0.1$                           & nm                  & \cite{Hall_1987} \\
$\gamma$    & $1.5$                             & ----                  & \cite{Moelans_2011} \\
$E_v^f$        &   $3.0$                           & eV                   & \cite{Yulan-Li_2013} \\
$E_g^f$        &   $4.668$                         & eV                   & \cite{Matthews_2020} \\ 
$c_v^{m,eq}$ &  $\mathrm{exp}\left(-E_{v}^f/k_{B}T\right)$ &---  & \cite{Larry_2019} \\
$c_g^{m,eq}$ &  $\mathrm{exp}\left(-E_{g}^f/k_{B}T\right)$ &--- & \cite{Larry_2019} \\
$c_v^{b,eq}$ &   $0.546$                       &---                     & \cite{Larry_2019} \\
$c_g^{b,eq}$ &   $0.454$                       &---      		      & \cite{Larry_2019} \\ 
$\mathcal{A}_{g}^{m}=\mathcal{A}_{v}^{m}$      &   $4.81\times10^{11}$         & J/{m$^3$}       & \cite{Larry_2019}  \\
$\mathcal{A}_{g}^{b}=\mathcal{A}_{v}^{b}$       &   $9.0\times10^{10}$               & J/{m$^3$}        & \cite{Larry_2019}      \\ 
$D_{g}^{m}=D_{v}^{m}$ & $D_{1} + D_{2} + D_{3}$ & m$^2$/s & \cite{Turnbull, pizzocri2018model}\\
 & $D_{1}= 7.6 \times 10^{-10}\mathrm{exp}\left(-4.86 \times 10^{-19}/k_{B}T\right)$ & & \\
 & $D_{2}= 4 \times 1.41 \times 10^{-25}\left[\sqrt{\dot{F}}\mathrm{exp}(-13800/T) \right] $ & & \\
 & $D_{3} = 2.0 \times 10^{-40}\dot{F}$ & & \\
 $D_{g}^{b}=D_{v}^{b}$  & $D_{g}^{m}$ & & \cite{Larry_2019,muntaha2023}\\
 $D_{g}^{gb}=D_{v}^{gb}$  & $1.3 \times 10^{-4} \mathrm{exp}(-32715.9/T)$ & m$^2$/s& \cite{muntaha2023}\\
 $D_{g}^{s}=D_{v}^{s}$  & $  5 \times 10^{-2} \mathrm{exp}(-36232/T)$ & m$^2$/s & \cite{muntaha2023}\\
\hline  
\end{tabular}
\end{center}
\label{table1}
\end{table}


Equation \eqref{DC} of the cluster dynamics model is solved using the finite difference method with implicit time integration, as implemented in the Xolotl simulation package~\cite{DongUk-Kim,muntaha2023}. Equations~\eqref{Eq:chem_g_evolution}, \eqref{Eq:AC1_evolution}, and \eqref{Eq:AC2_evolution} of the PF model are solved using the finite element method, as implemented in the MARMOT~\cite{DongUk-Kim,muntaha2023} code based on the MOOSE framework \cite{permann2020moose}. The discretized PF equations are solved using the preconditioned Jacobian-free Newton-Krylov method. A second-order backward differentiation scheme is used to evolve the system in time. To reduce computational costs, an adaptive time-stepping scheme is used, as described in our previous work ~\cite{muntaha2023}. For all simulations, the linear convergence tolerance is set at $10^{-5}$ and the non-linear absolute and relative tolerances are $10^{-8}$ and $10^{-9}$, respectively.

The two codes are coupled using a weak coupling scheme where, at the start of each time step, Xolotl passes gas arrival rates to MARMOT at GBs and bubble surfaces, and MARMOT returns the evolving interface locations to Xolotl. Data exchange is managed through the MOOSE framework using its MultiApp system \cite{permann2020moose}, with MARMOT serving as the master application and Xolotl as a sub-application connected via an intermediate finite element wrapper. This approach allows each code to maintain its native numerical methods, finite difference for Xolotl and finite element for MARMOT, while enabling seamless information transfer between the grain-scale cluster dynamics and interface-scale PF calculations. For more detailed information about how the codes are coupled, readers are referred to Kim \emph{et al.}~\cite{DongUk-Kim}.

\subsection{Simulation setup and boundary conditions}
\label{setup_details}

To investigate the impact of GB and intergranular bubble interactions on FGR, we use the hybrid model to simulate the microstructure evolution in 3D UO$_2$ polycrystals. We simulate the behavior that would occur in a fuel rod with a fission rate $\dot{F} = 3.27\times10^{19}$ fissions/(m$^3$s). UO$_2$ fuel pellets exhibit large temperature gradients from the fuel centerline to the fuel FS, ranging from around 700 K at the outer edge to around 1700 K or higher in the pellet center \cite{bagger1994temperature}. FGR due to intergranular bubble coalescence has been shown to start around 1500 K \cite{bagger1994temperature}. Therefore, we simulate the microstructure evolution at 1200 K, representing an intermediate location in the fuel pellet where we would not expect significant FGR, and 1600 K, representing a location near the center where FGR should occur.

For our simulations, we generate two polycrystal UO$_2$ microstructures: one with 10 initial grains and another with 100. The grain structures are generated using MicroStructPy~\cite{hart2020microstructpy}, a microstructure generator software package. Subsequently, the grain centroid locations are imported from MicroStructPy into MARMOT to serve as the center locations of 3D Voronoi tessellations. To ensure that the mean grain size is around $10$ $\mu$m, a typical value for UO$_2$ fuel pellets in commercial LWRs \cite{Larry_2019}, we use a $20\times20\times20\ \mu$m$^3$ domain for the 10-grain polycrystal and $40\times40\times40\ \mu$m$^{3}$ for the 100-grain polycrystal. The initial mean grain radii in the 10- and 100-grain polycrystals are $11.6$ and $10.6$ $\mu$m, respectively. The computational cost of a PF simulation increases as the interfacial width $l_w$ decreases, since a finer mesh re-solution is required to resolve the diffuse interface, so we use a larger interfacial width in the 100-grain polycrystal simulations to reduce the cost: $l_w$ is 480 nm and 960 nm in the 10- and 100-grain simulations, respectively (Table \ref{table1}).

\begin{figure}[tb]
 \centering
 \includegraphics[width=0.9\linewidth]{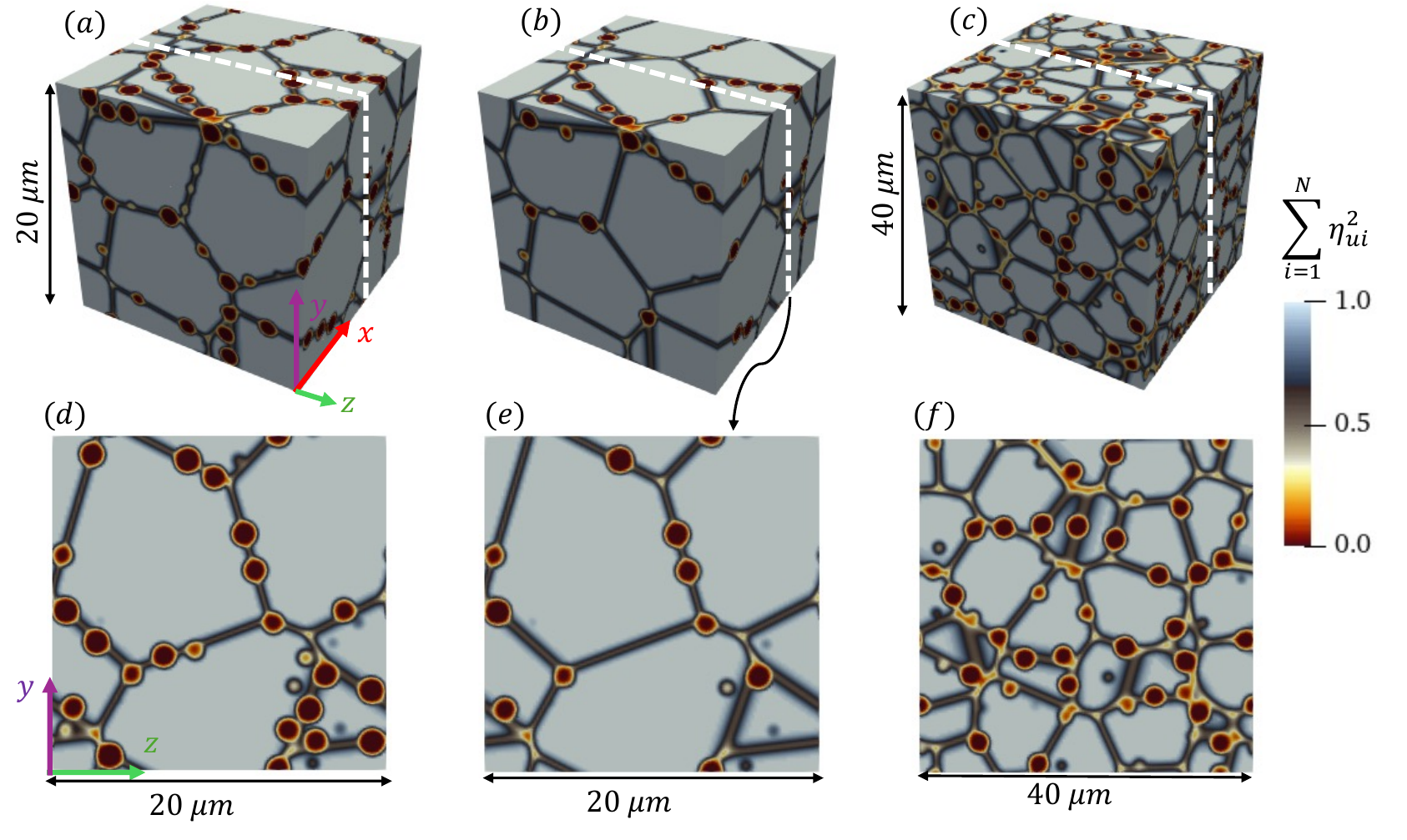}
 \caption{The initial microstructures with periodic boundary conditions, where (a) and (b) show the 10-grain polycrystals with 320 and 160 intergranular bubbles, respectively, and (c) shows the $100$-grain polycrystal with 600 bubbles. (d)--(f) show 2D slices of the 3D microstructures, indicated by the white dotted lines in (a)--(c). The images are shaded by the function  $\Psi=\sum_{i=1}^{N}\eta_{ui}^2$, with light grey ($\Psi \approx 1$) representing grains, dark grey ($\Psi \approx 0.5$) representing GBs and red-yellow ($\Psi<0.4$) representing bubbles. }
\label{Fig1}
\end{figure}

The focus of our work is to understand the interactions between intergranular bubbles and GBs, therefore, we do not consider the nucleation of intergranular bubbles. Rather, we start the simulation with initial bubbles that are the smallest and are stable in the PF model. To be stable, the radius of the bubble must be $\geq1.2\,l_w$. Therefore, the initial bubble radii in the $10$- and $100$-grain cases are $710$ and $1140$ nm, respectively. The bubbles are randomly positioned on the GBs. To avoid bubble overlap between initial bubbles, the minimum spacing between the bubble centers is set to $3.5\,l_w$ for all cases. To understand the impact of initial bubble density on bubble evolution, we consider initial 10-grain polycrystals with 320 and 160 bubbles, as shown in Figs.~\ref{Fig1}a-b, with initial GB bubble densities of $0.15$ and $0.075$ $\mu$m$^{-2}$, respectively. The 100-grain polycrystal has $600$ initial bubbles, as shown in Fig.~\ref{Fig1}c, has an initial GB bubble density of $0.032$ $\mu$m$^{-2}$. In these initial grain structures, the smaller the bubble density, the larger the average distance between bubbles.

In the 10-grain simulations, we use $10$ order parameters to represent UO$_2$ grains. However, using 100 order parameters in the $100$-grain simulations would have a prohibitively large computational cost. Therefore, we employ $25$ order parameters to represent the $100$ grains such that each order parameter represents multiple grains. To avoid coalescence when two grains represented by the same order parameter come in contact, we use the grain tracker algorithm~\cite{PERMANN201618} from the MOOSE framework to remap one of the grains to another order parameter. 

All simulations are performed using uniform meshes of linear hexahedral finite elements. Similar to Aagesen \emph{et al.}~\cite{Larry_2019}, the element size $\Delta x$ is assumed to be one-third the interface width $l_{w}$ to adequately resolve the diffuse interface. This results in $125\times125\times125$ elements for both the 10- and 100-grain simulations. The mesh is divided across multiple computer processors to reduce the memory usage associated with outputting and visualizing the results.

We carry out simulations of polycrystals that are away from and touching a FS. For the polycrystals without a FS, we use periodic boundary conditions for all variables at all boundaries. For the polycrystals with a FS, the FS is the side of the domain at $x=0$. The boundary conditions at the FS and the side opposite the FS are the same as those used in Muntaha \emph{et al.}~\cite{muntaha2023}: for the FS, zero flux for the grain order parameters and Dirichlet boundaries with $\eta_{b0}=0$, $\mu_v=0$, and $\mu_g=-0.023$; for the opposite side, zero flux for all variables. The negative chemical potential boundary condition for the gas atoms at the FS was fit to ensure that the gas cannot enter through the FS. Note that the initial 3D 10- and 100-grain microstructures with a FS differ marginally from the periodic microstructures depicted in Fig.~\ref{Fig1} in the direction parallel to the FS normal. This variation arises because the initial grain structure created by the Voronoi tessellation is not periodic in that direction.

The size of these 3D polycrystal simulations using the hybrid model is very large. The number of non-linear degrees of freedom (DOFs) for the 10-grain simulations in MARMOT and Xolotl are $26\times 10^6$ and $2\times10^9$, respectively. The latter was calculated by multiplying the total number of grid points by the number of Xe clusters in the network. These simulations were run on $960$ cores on the Sawtooth cluster at the Idaho National Laboratory. In the 100-grain simulations, the DOFs in MARMOT and Xolotl are $56\times10^6$ and $2\times10^9$, respectively. They were run on $2400$ cores on the Sawtooth cluster. The simulated time depends on the simulation and is determined by the overall wall time taken to run the simulation. Table~\ref{table2} summarizes the computation time and end time for the 10- and 100-grain simulations with and without a FS. To the best of our knowledge, these are the first large-scale simulations employed to predict the evolution of fission gas bubbles in UO$_2$ polycrystals.

\begin{table}[btp]
\captionsetup{justification=raggedright, singlelinecheck=false}
\caption{Summary of all the simulations carried out in this work, including the initial number of grains, whether they have a FS, the temperature, the initial number of bubbles, the computation time, and the end time.}
\begin{center}
\begin{tabular}{m{1.9cm}ccm{1.9cm}m{1.9cm}c}
\hline
\centering Initial no. of grains & FS & Temperature [K] & \centering Initial no. of bubbles & \centering Computation time [days] & End time [days] \\
\hline
\centering \multirow{4}{*}{10} &  \centering \multirow{3}{*}{No}    & $1200$    & \centering $320$    & \centering $3.0$     & $515.0$   \\
 					      						       && $1600$    & \centering $320$    &\centering $3.1$      & $131.8$   \\
 					      						       && $1600$    & \centering $160$    & \centering$4.6$      & $57.9$\\
					    & Yes 					         & $1600$    & \centering $300$     &\centering$1.0$       & $155.9$\\[1ex]
\centering \multirow{3}{*}{100} &  \centering \multirow{2}{*}{No}  & $1200$    &\centering $600$      & \centering $3.8$      & $224.8$   \\
 					       							&& $1600$   &\centering $600$     &\centering $30.1$     & $389.6$   \\
						& Yes					 & $1600$    &\centering $600$     &\centering $6.7$       & $320.4$ \\	
\hline
\end{tabular}
\end{center}
\label{table2}
\end{table}

%% file: Results.tex
\section{Results}
\label{sec:results}
In this section, we present the results of our large 3D simulations of fission gas bubble evolution in UO$_2$ polycrystals. We begin by comparing the results of our simulation at 1600 K of the 10-grain polycrystal with 320 initial bubbles with experimental data from White~\cite{White_2004}, to provide some validation of the hybrid model. We then present the results of the 10-grain simulations without a FS, followed by  100-grain simulations with same boundary condition. We end with the results from both the 10- and 100-grain simulations with a FS. 

To make quantitative comparisons, we track the number of bubbles and grains as a function of time in all simulations. In addition, we calculate GB and TJ coverages over time, as detailed in the Appendix~\ref{AppendixA2}. 

\subsection{Validation of bubble coalescence behavior}
\label{sec:results_validation}
We begin by comparing the intergranular bubble coalescence behavior in the simulation without a FS with 10 grains and 320 bubbles at 1600 K with the experimental data for test \#4004 from White (see Table 1 from \cite{White_2004}). White characterized the intergranular bubbles from the sample that reached a moderate burnup of 20.5 GWd/tU. The relationship between intergranular bubble density $N$, defined as the number of bubbles per unit GB area, and the mean intergranular bubble area $A$, defined as the area of the bubbles projected on the GB plane, was characterized for 8010 bubbles on 44 GB planes. Millet \emph{et al.}~\cite{Millet_2012_a} and Prudil \emph{et al.}~\cite{PRUDIL2022153777} both compared their 3D simulation results to these data and showed reasonable agreement. To compare our prediction with their results, we extracted their simulation data using WebPlotDigitizer \cite{WPD}.

\begin{figure}
    \centering
    \includegraphics[trim=0 0 0 0, clip, keepaspectratio,width=0.51\linewidth]{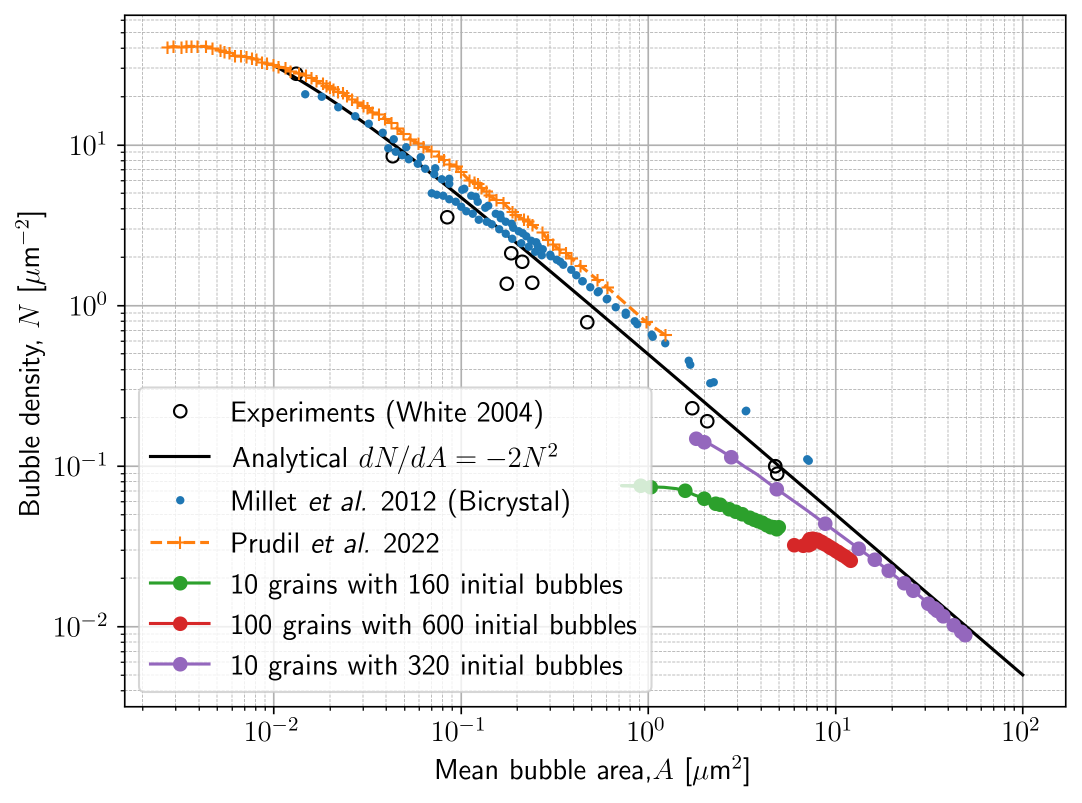}
    \caption{Comparison of our simulation results with experimental data and an analytical model (Eq.~\eqref{Eq:White_Analytical}) from White \cite{White_2004} for the bubble density $N$ versus mean bubble area $A$. Our simulation results without a FS at 1600 K are show for the 10-grain polycrystal with 320 and 160 initial bubbles and for the 100-grain polycrystal. We include the simulation data from Millet \textit{et al.}~\cite{Millet_2012_a} and Prudil \textit{et al.}~\cite{PRUDIL2022153777} for reference.}
    \label{fig:validation}
\end{figure}

Figure~\ref{fig:validation} shows the variation in the simulated bubble density with mean projected bubble area from our simulation results compared with the experimental data. These data show that as the mean bubble area on a GB increases, the bubble number density decreases due to coalescence and coarsening. White \cite{White_2004} proposed the following relationship between bubble density (N) and mean bubble area (A) due to coalescence:
\begin{equation}
    \frac{dN}{dA}=-2N{^2}. 
    \label{Eq:White_Analytical}
\end{equation}
In Fig.~\ref{fig:validation}, we plot the analytical curve along with the data. White~\cite{White_2004} noted that the majority of experimental data remain to the left of the analytical curve because coalescence begins once the bubble area exceeds this curve. Following coalescence, the bubble area shrinks to reduce the total bubble surface area. 

Our simulation results consistently fall to the left of the analytical curve, similarly to the experimental data. This is distinct from the results of Millet \emph{et al.}~\cite{Millet_2012_a} and Prudil \emph{et al.}~\cite{PRUDIL2022153777}, which both fall above the analytical line. This is potentially due to the incorporation of GB migration in our simulations, which was absent in the other models, which could accelerate coalescence. Also note that our simulations reach smaller bubble densities and larger mean bubble areas than the other simulations and the data. This is due to our large initial bubble radius required by our mesh size, which we use to reduce the computational cost of these large simulations. Figure~\ref{fig:validation} also shows the curves from the 10-grain polycrystal with 160 initial bubbles and the 100-grain polycrystal at 1600 K without a FS. In these simulations, the initial bubbles are further apart with lower bubble initial density and do not experience significant coalescence, thus they are far to the left of the analytical curve, as shown in Fig.~\ref{fig:validation}.

\subsection{10-grain polycrystals without a FS}
\label{sec:10-grain}

\begin{figure}[tbp]
 \centering
 \begin{subfigure}{0.9\linewidth}
   \includegraphics[width=\linewidth]{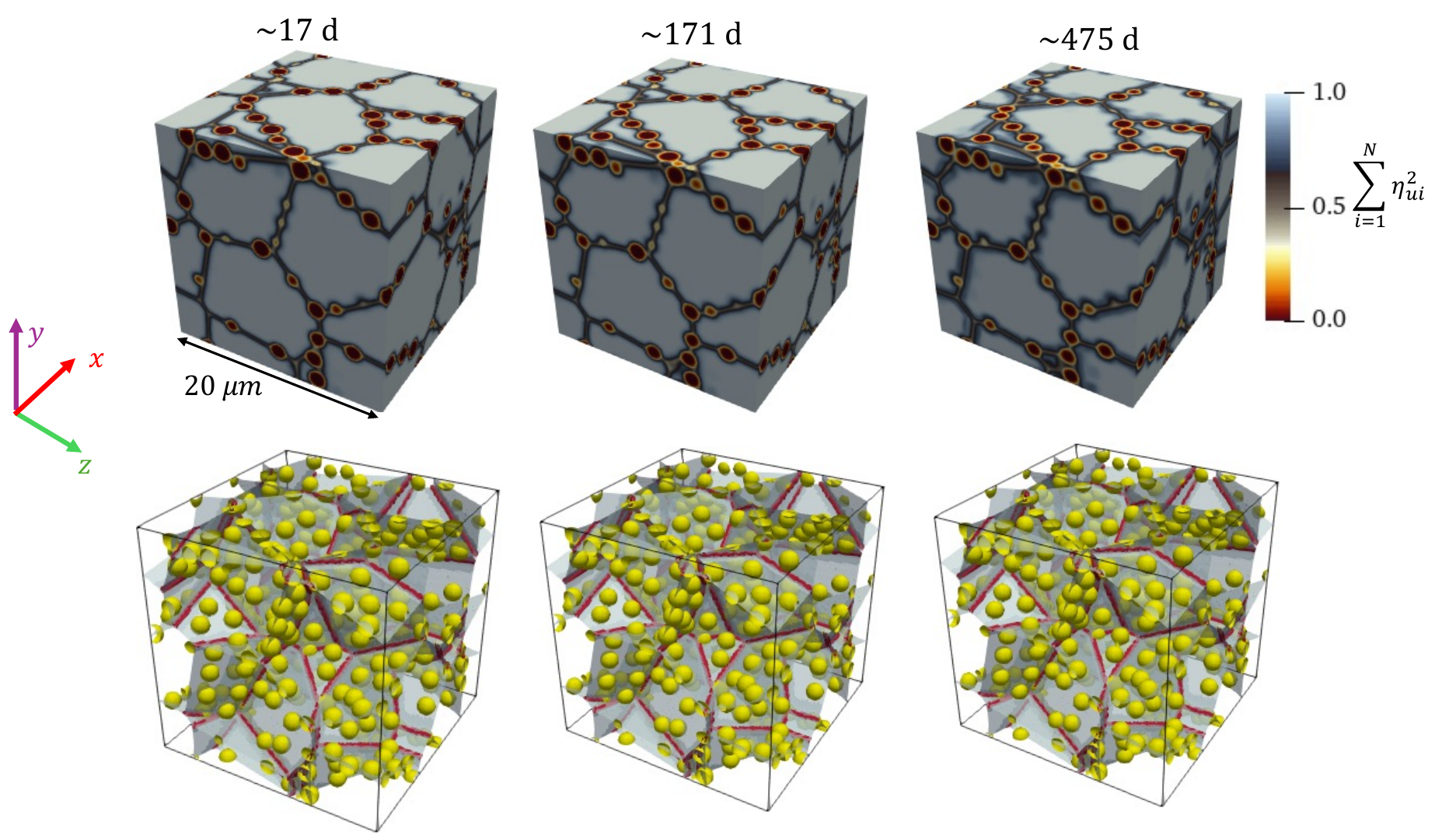}
   \caption{}\label{fig:evol_10gr_320_1200K}
 \end{subfigure}
 \begin{subfigure}{0.9\linewidth}
     \includegraphics[width=\linewidth]{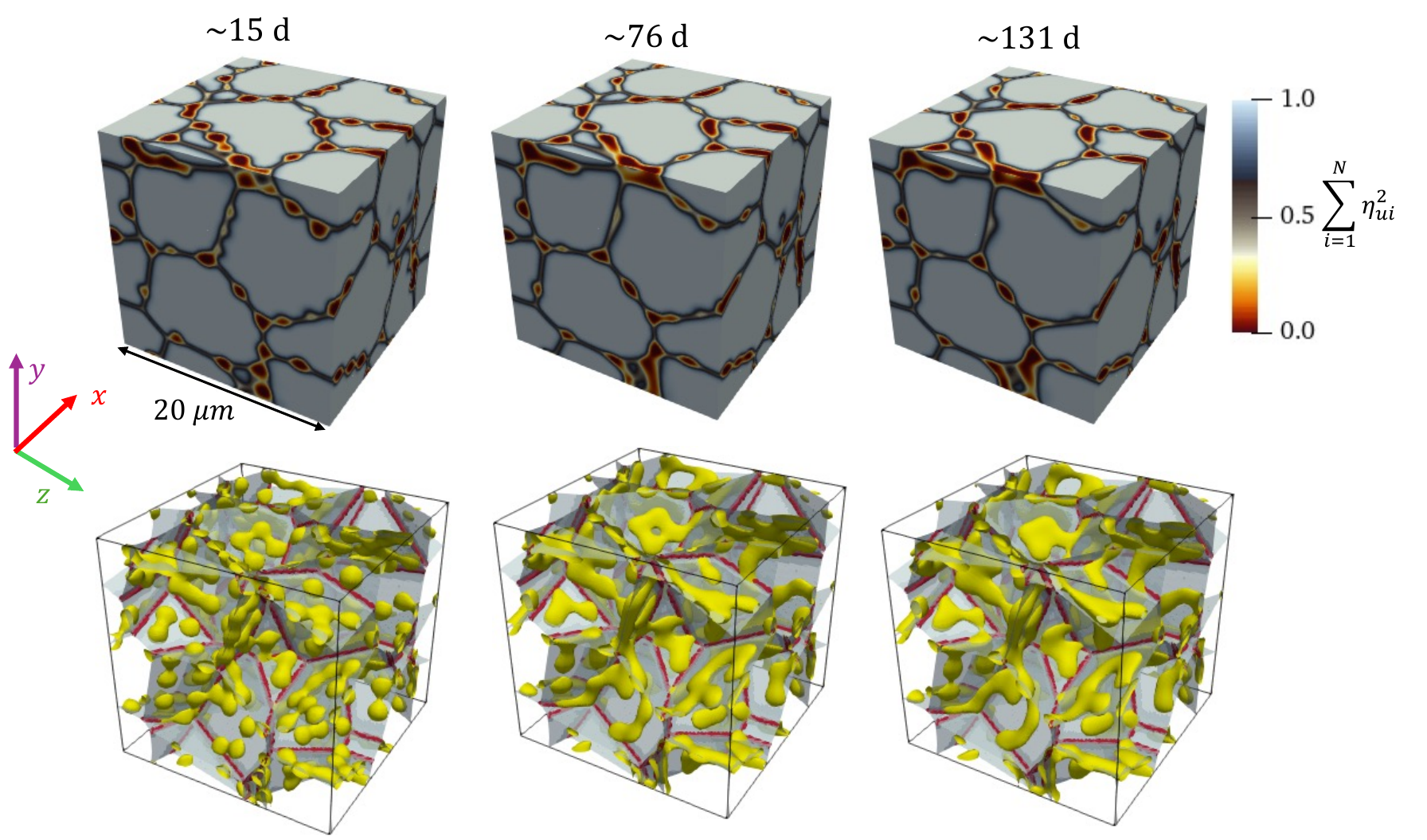}
     \caption{}\label{fig:evol_10gr_320_1600K}
 \end{subfigure}
 \caption{Evolution of 10-grain UO$_2$ polycrystals without a FS with $320$ initial bubbles at (a)  1200 K and (b) 1600 K. The top images are shaded by the function  $\Psi=\sum_{i=1}^{N}\eta_{ui}^2$, with light gray ($\Psi \approx 1$) representing grains, dark gray ($\Psi \approx 0.5$) representing GBs and red-yellow ($\Psi<0.4$) representing bubbles. In the bottom images, the intergranular bubble contour ($\eta_{b0}=0.5$) is yellow, the initial GBs without bubbles are light gray, and the initial TJ lines without bubbles are red.}
\label{fig:evol_10gr_320}
\end{figure}

\begin{figure}[tbp]
\centering
\begin{subfigure}{0.49\textwidth}
	\includegraphics[trim=0 0 0 0, clip, keepaspectratio,width=\linewidth]{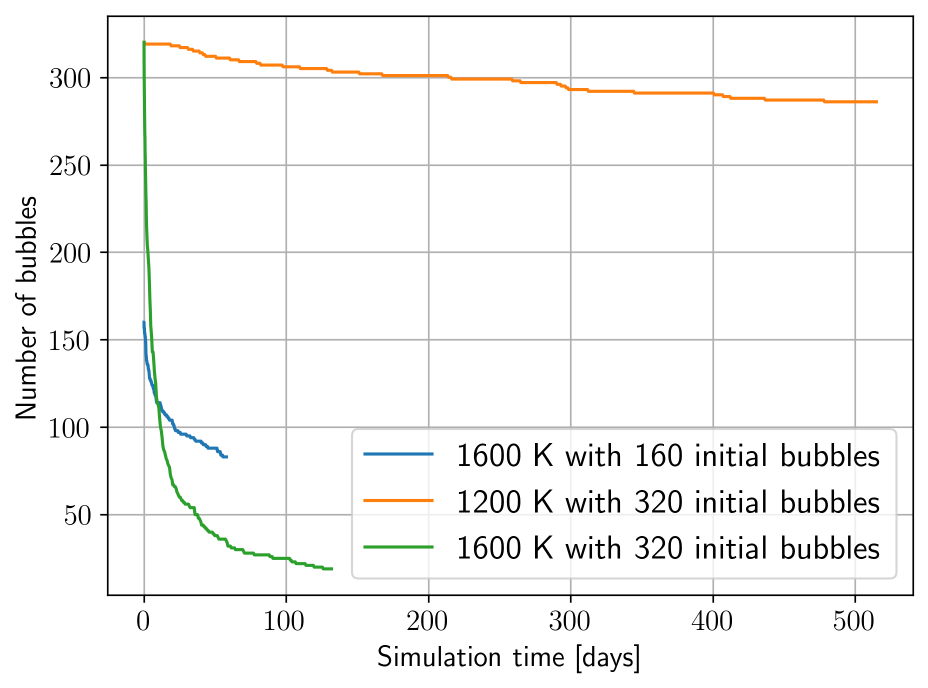}
	\subcaption{}
	\label{Fig4a}
\end{subfigure}
\begin{subfigure}{0.49\textwidth}
	\includegraphics[trim=0 0 0 0, clip, keepaspectratio,width=\linewidth]{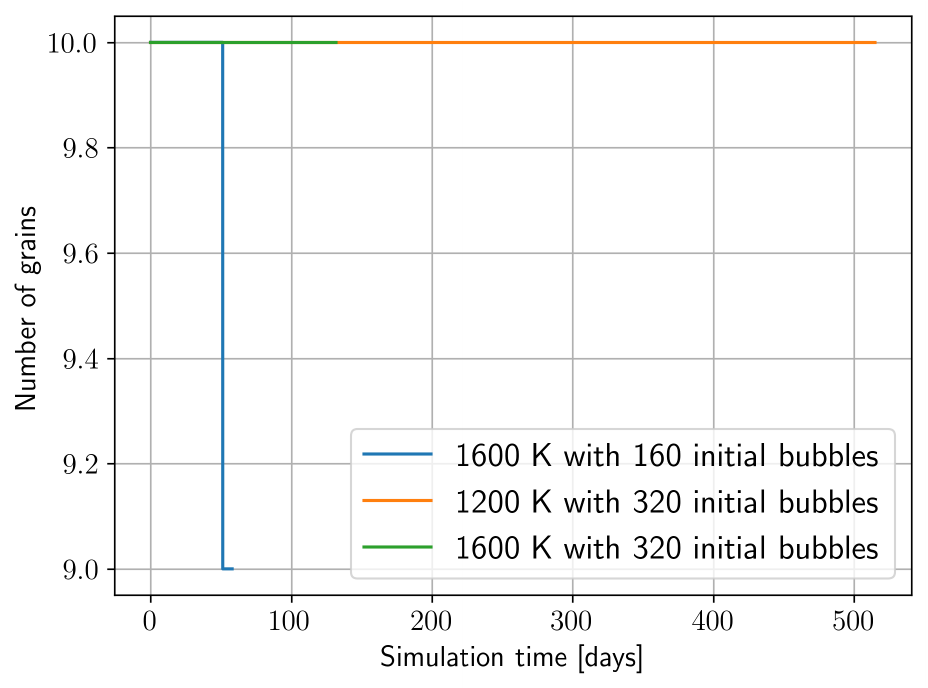}
	\subcaption{}
	\label{Fig4b}
\end{subfigure}
\begin{subfigure}{0.49\textwidth}
	\includegraphics[trim=0 0 0 0, clip, keepaspectratio,width=\linewidth]{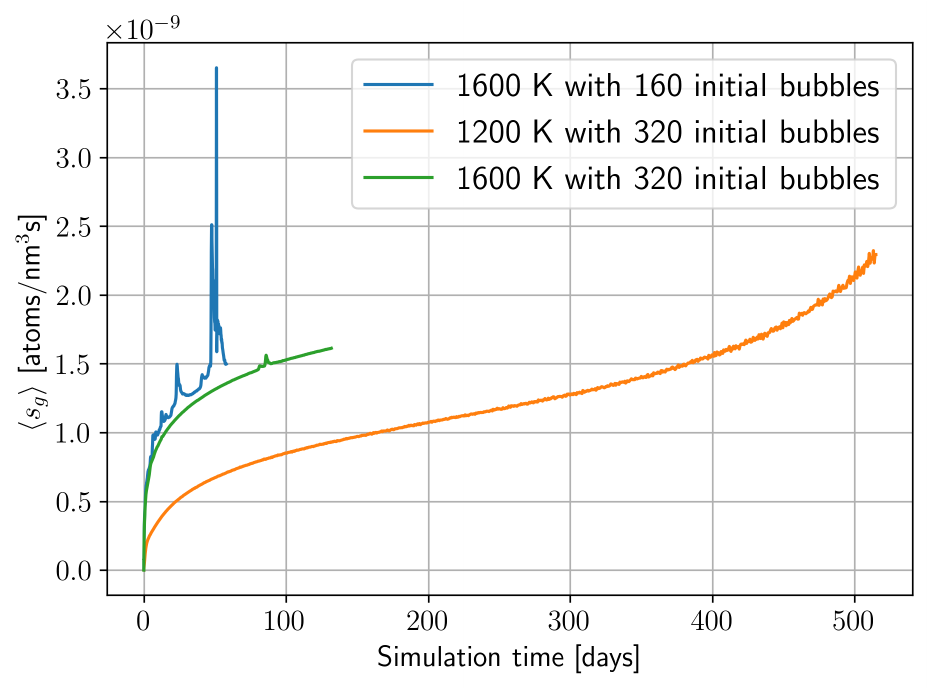}
	\subcaption{}
	\label{Fig4c}
\end{subfigure}
\begin{subfigure}{0.49\textwidth}
	\includegraphics[trim=0 0 0 0, clip, keepaspectratio,width=\linewidth]{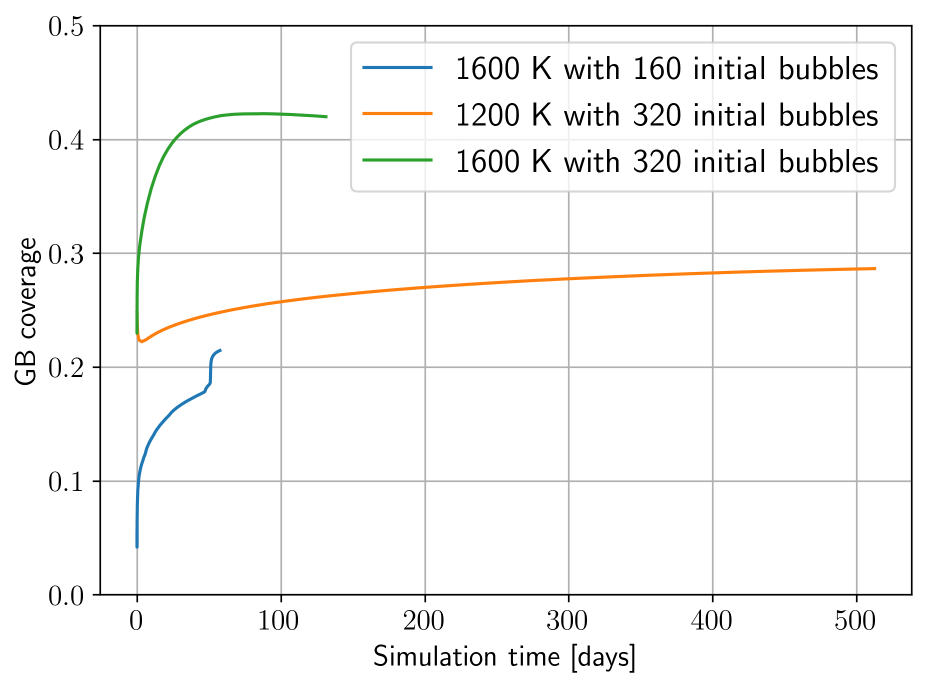}
	\subcaption{}
	\label{Fig4d}
\end{subfigure}
\begin{subfigure}{0.49\textwidth}
	\includegraphics[trim=0 0 0 0, clip, keepaspectratio,width=\linewidth]{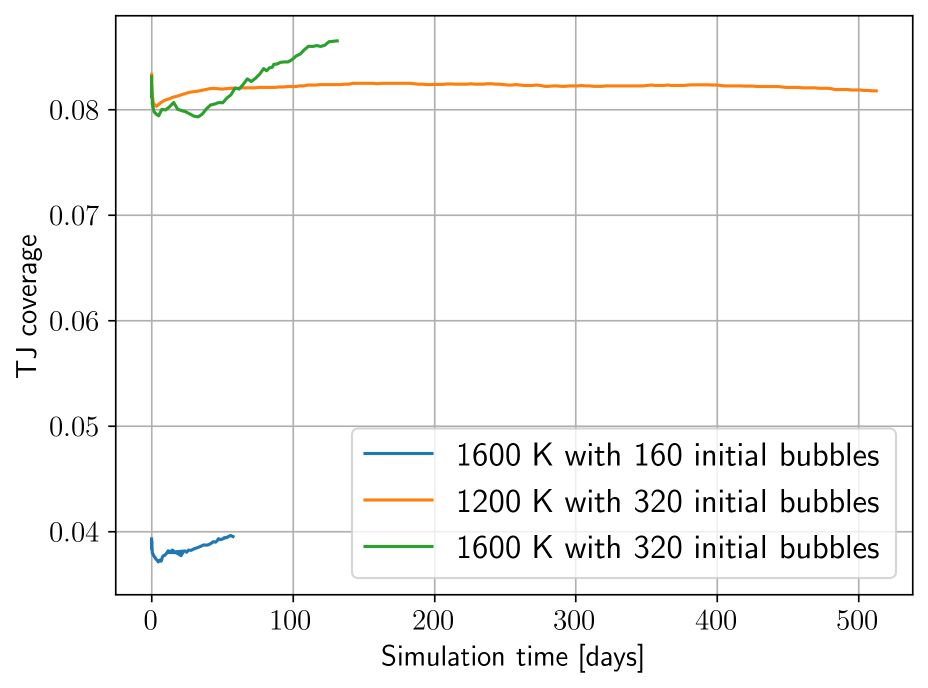}
	\subcaption{}
	\label{Fig4e}
\end{subfigure}
%
%
\caption{Metrics evolution with time quantifying the microstructure evolution in the $10$-grain polycrystals without a FS, where (a) shows the number of intergranular bubbles, (b) number of grains, (c) $\langle s_{g}\rangle$, representing the volume-averaged number of Xe atoms reaching GBs and intergranular bubbles surfaces, calculated using Eq.~\eqref{eq:av_sg}, (d) GB fractional coverage, and (e) TJ fractional coverage. }
\label{Figure4}
\end{figure}

We begin with a comparison of the microstructure evolution at 1200 K and 1600 K in the 10-grain polycrystals with 320 initial bubbles and no FS. Images of the evolving microstructures are shown in Fig.~\ref{fig:evol_10gr_320}. We visualize the GBs and bubbles on the outer boundaries of the 3D domain and we show the evolution of the bubbles throughout the domain. Metrics quantifying the microstructure evolution are shown in Fig.~\ref{Figure4}.

At 1200 K, we observe no significant evolution with time for either gas bubbles or GB planes, as shown in Fig.~\ref{fig:evol_10gr_320_1200K}. The bubble count decreases by roughly $10$\%, from $320$ to $286$, over $\sim500$ days (Fig.~\ref{Fig4a}) which is likely due to the coalescence of bubbles close enough to touch as they evolve from a spherical to a lenticular shape. No change is observed in the grain count (Fig.~\ref{Fig4b}). This lack of evolution is because both the Xe gas diffusivity and UO$_2$ GB mobility are small at $1200$ K. Note that the bubble evolution behavior in this study differs from that reported by Lan~\textit{et al.}~\cite{lan2024three}, where they observed extensive bubble coalescence at $1273$ K, despite including only bulk gas diffusivity. This is likely due to their model not including trapping of gas atoms by intragranular bubbles, which is captured in our hybrid model by Xolotl.

At 1600 K, significant intergranular bubble evolution is observed, as shown in Fig.~\ref{fig:evol_10gr_320_1600K}. Intergranular bubbles initially shift from being spherical to being lenticular, and eventually coalesce to form interconnected networks. We can also observe bubble coalescence at TJs. This morphological transition agrees with experiments~\cite{White_2004} and previous 3D simulations~\cite{Larry_2019, lan2024three}. The bubble count decreases by $\sim94$\%, from 320 to 19, over 131 days (Fig.~\ref{Fig4a}). No significant GB migration is observed for any of these cases, since they are pinned by the intergranular bubbles. As a result, the number of grains remains constant with time (Fig.~\ref{Fig4b}). 

To quantify the amount of gas moving from the grains to the GBs and intergranular bubbles surfaces, we use the Xe source term $s_g$ that passes gas atoms from Xolotl to MARMOT. Specifically, we calculate the volumetric average of the source term as a function of time:
\begin{equation}
    \langle s_{g}\rangle = 1/V\int_{V}s_{g}dv.
    \label{eq:av_sg}
\end{equation}
This quantity results from the diffusion of single Xe atoms to the GBs and sweeping of immobile Xe clusters by migrating GBs~\cite{DongUk-Kim}. Figure~\ref{Fig4c} shows that the rate of arrival of gas at GBs and bubble surfaces increases with time at both $1200$ and $1600$ K. This is likely a result of increasing re-solution as the concentration of gas trapped in intragranular bubbles increases. The arrival rate at $1600$ K is almost twice that at $1200$ K after $100$ days, even though the gas production rate within the grains is the same. This is because Xe atoms diffuse faster within the UO$_2$ grains at $1600$ K and make it to the GB and intergranular bubble surfaces, unlike the case at 1200 K.

Analytical FGR models use GB coverage, defined as the ratio of the GB area covered by bubbles to the total GB area,  as a criterion for the initiation of FGR  \cite{Pastore_2013}; a GB coverage of 50\% is often used to determine the initiation. Several 3D mesoscale fission gas studies \cite{Millet_2012_a, Larry_2019, lan2024three} have reported the evolution of GB coverage, but never for complex evolving grain structures. In this study, we adopt the GB coverage calculation from Aagesen \emph{et al.}~\cite{Larry_2019}, and extend it to polycrystalline UO$_2$. At $1200$ K, the GB coverage increases from the 25\% of the initial structure to a little over 30\% in $\sim500$ days, as shown in Fig.~\ref{Fig4d}, indicating that the intergranular bubble coalescence is negligible. This  amount of interconnection is well below the 50\% threshold often assumed for FGR, and so our result is consistent with the experimental finding that the FGR temperature threshold is 1500 K \cite{bagger1994temperature}. At 1600 K, the GB coverage increases sharply, from $\sim25$\% to $43$\% in $\sim50$ days, and then saturates once all bubbles coalesce. The saturation value remains below the critical 50\% used in  engineering-scale FGR models \cite{Pastore_2013}. We will directly model FGR for these conditions in Section~\ref{sec:results_FS}. 

TJ tunnels are assumed to be a major contributor in Stage 3 of FGR. So, we also calculate TJ coverage, defined as the ratio of the total TJ length occupied by bubbles to the total TJ length, as a function of time \cite{Larry_2019}. As noted in Appendix~\ref{AppendixA2}, we assume that the total TJ length without bubbles remains constant with time and is calculated based on the initial microstructure. This assumption remains valid, provided that grain growth is negligible, as observed in our 10-grain simulations.

Figure~\ref{Fig4e} shows that TJ coverage at $1200$ K shows negligible variation with time, but at $1600$ K, it marginally increases (from 8\% to around 8.7\%). Note that our initial TJ coverage ($\sim8\%$) is significantly lower than the values reported in Aagesen~\textit{et al.}~\cite{Larry_2019} ($>$$50$\%) and Lan~\textit{et al.}~\cite{lan2024three} ($\sim40$\%). This is because, in these studies \cite{Larry_2019, lan2024three}, bubbles were purposefully placed along the TJ lines at time $t=0$. In contrast, our simulations were initialized by randomly placing bubble centers at the GBs and not at the TJs, resulting in significantly lower initial TJ coverage. Since the TJ coverage only increases by less than 1\% after 120 days at $1600$ K, this suggests that the GB bubbles do not migrate to cover the TJ lines in our simulations.



%

In their $3$D bicrystal microstructures, Millet \textit{et al.}~\cite{Millet_2012_a} found that a twofold increase in the initial GB bubble density resulted in a threefold decrease in the time required to reach the critical GB coverage of $50\%$. However, their model assumed homogeneous gas diffusion and neglected the production, trapping, and re-solution of fission gas atoms within the grains. We investigate the impact of the initial intergranular bubble density on GB bubble evolution in our hybrid simulations by performing a 10-grain polycrystal simulation at 1600 K with 160 initial intergranular bubbles rather than 320; the initial microstructure is shown in  Fig.~\ref{Fig1}b.

\begin{figure}[tbp]
 \centering
 \begin{subfigure}{0.9\linewidth}
    \includegraphics[width=\linewidth]{Figure3.pdf}
    \caption{}
    \label{fig:10gr_1600K_320b}
 \end{subfigure}
 \begin{subfigure}{0.9\linewidth}
    \includegraphics[width=\linewidth]{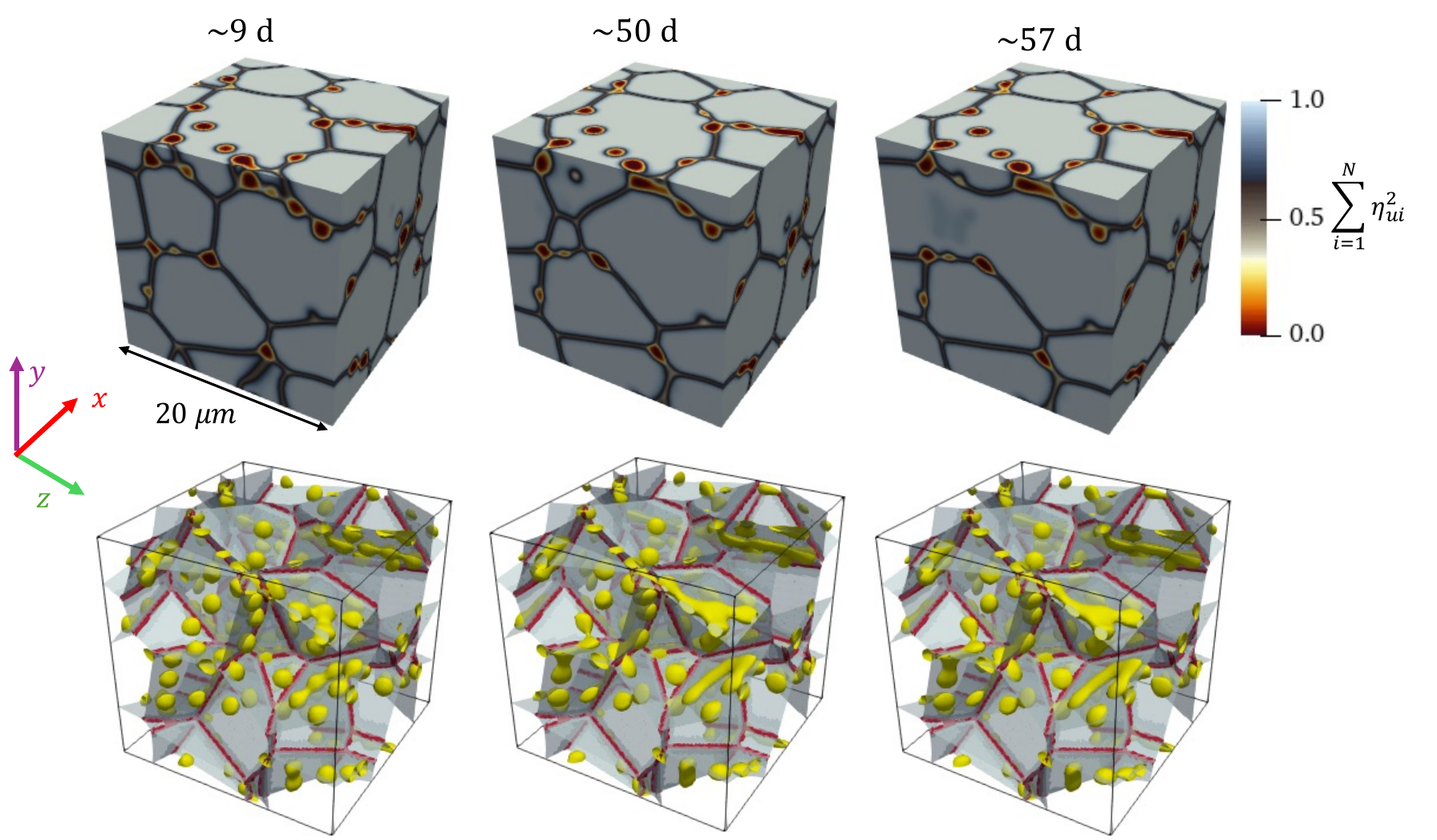}
    \caption{}\label{fig:10gr_1600K_160b} 
 \end{subfigure}
 \caption{Evolution with time of 10-grain UO$_2$ polycrystals at 1600 with (a) 320 initial bubbles (same as Fig.~\ref{fig:evol_10gr_320_1600K}) and (b)  160 initial bubbles. The top images are shaded by the function  $\Psi=\sum_{i=1}^{N}\eta_{ui}^2$, with light gray ($\Psi \approx 1$) representing grains, dark gray ($\Psi \approx 0.5$) representing GBs and red-yellow ($\Psi<0.4$) representing bubbles.In the bottom images, the intergranular bubble contour ($\eta_{b0}=0.5$) is yellow, the initial GBs without bubbles are light gray, and the initial TJ lines without bubbles are red.}
\label{fig:10gr_1600K}
\end{figure}
Figure~\ref{fig:10gr_1600K} shows the evolving microstructures from the 320 (repeated from Fig.~\ref{fig:evol_10gr_320_1600K}) and 160 initial bubble simulations. The fewer initial bubbles results in larger average spacing between bubbles. This results in less bubble coalescence as indicated by the smaller reduction in bubble count with $160$ initial bubbles than with the $320$ initial bubbles (Fig.~\ref{Fig4a}); the bubble count decreases over 50 days by approximately $45$\% with 160 initial bubbles and  $88\%$ with $320$ initial bubbles. Fewer initial bubbles also result in reduced GB pinning, so we observe significant GB evolution that eventually results in the disappearance of a grain; the grain count decreases from $10$ to $9$ after $50$ days with 160 initial bubbles (Fig.~\ref{Fig4b}). It is noteworthy that the simulation time in the 10-grain cases decreases with decreasing initial bubble density, even though the wall times are similar (Table~\ref{table2}). This is because fewer intergranular bubbles cause more grain growth, leading to smaller time steps.

The GB migration directly impacts the arrival rate of gas atoms to the GBs and bubble surfaces. Figure \ref{Fig4c} shows that the average source term $\langle s_g \rangle$ has large spikes in it over time for the 160 initial bubble case, while the 320 bubble case is fairly smooth. The largest spike occurs slightly before the number of grains goes down from 10 to 9. This indicates an increase in the amount of gas atoms transferred to the UO$_2$ GBs due to the sweeping up of immobile Xe clusters by the migrating GBs. This demonstrates the importance of including intragranular bubbles in our hybrid model.

The initial GB coverage is much smaller in the 160 initial bubble simulation ($\sim5$\%) compared to the 320 bubble simulation ($\sim22$\%), as shown in Fig.~\ref{Fig4d}, though the change in the coverage over 50 days is similar in the two simulations at 1600 K. However, the GB coverage curve has not saturated in the 160 bubble simulation, while it has in the 320 bubble simulation. The TJ coverage increases marginally from $3.8\%$  to $3.9\%$ in the 160 bubble simulation (Fig.~\ref{Fig4e}), which is around a quarter of the increase in the 320 bubble simulation. As discussed previously for the 320 bubble simulation, this is because our simulations start with almost no TJ bubbles, and the GB bubble network does not significantly overlap with the TJ lines. However, it is important to note that our method used to calculate the TJ coverage (described in Appendix~\ref{AppendixA2}), uses the initial TJ lines and so will decrease in accuracy with time in this case since some GB migration occurs.

\subsection{100-grain polycrystals without a FS}
\label{Section3.4}

\begin{figure}[p]
 \centering
 \begin{subfigure}{0.9\linewidth}
     \includegraphics[width=\linewidth]{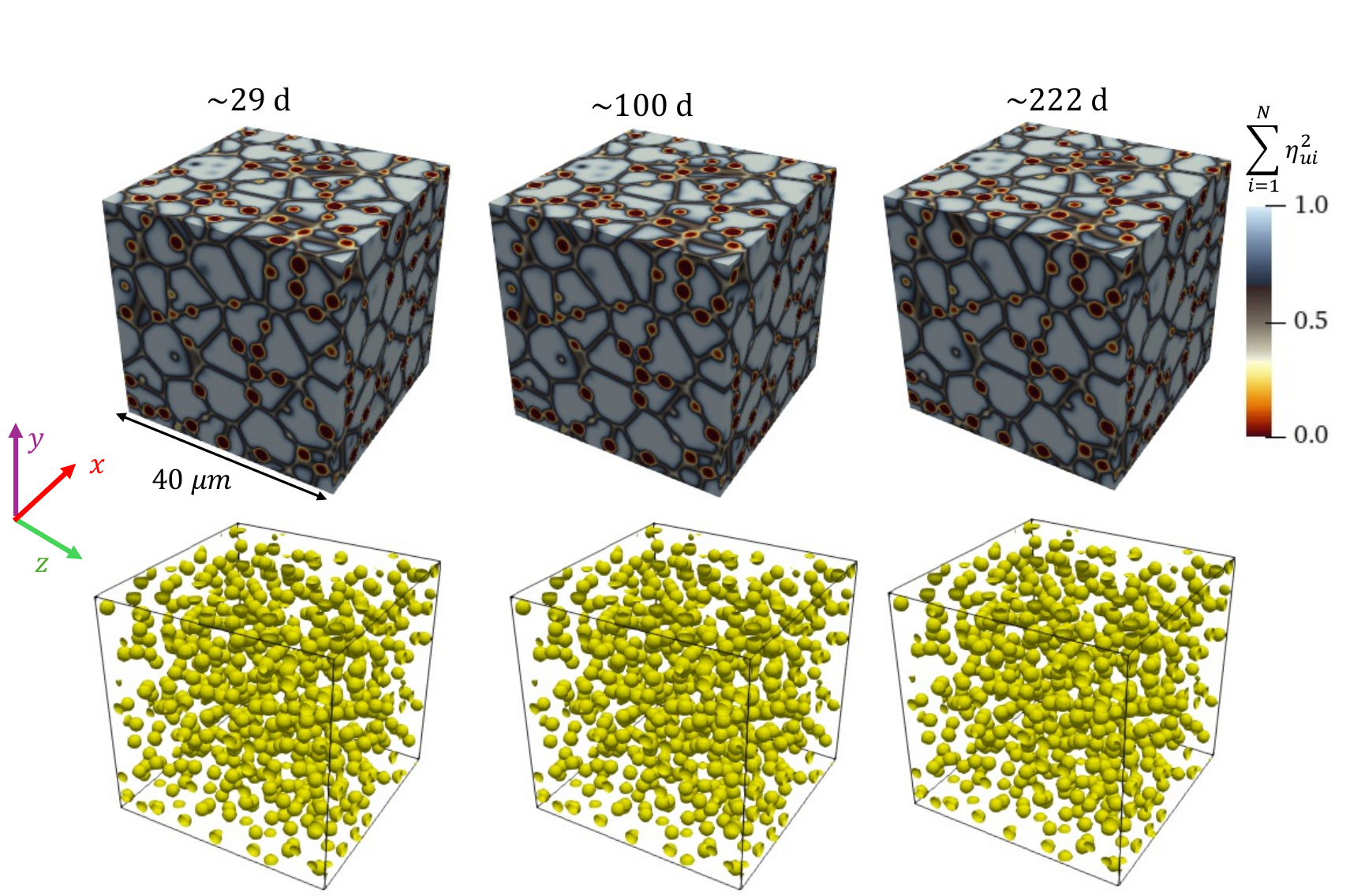}
     \caption{} \label{fig:evol_100gr_1200K}
 \end{subfigure}
 \begin{subfigure}{0.9\linewidth}
     \includegraphics[width=\linewidth]{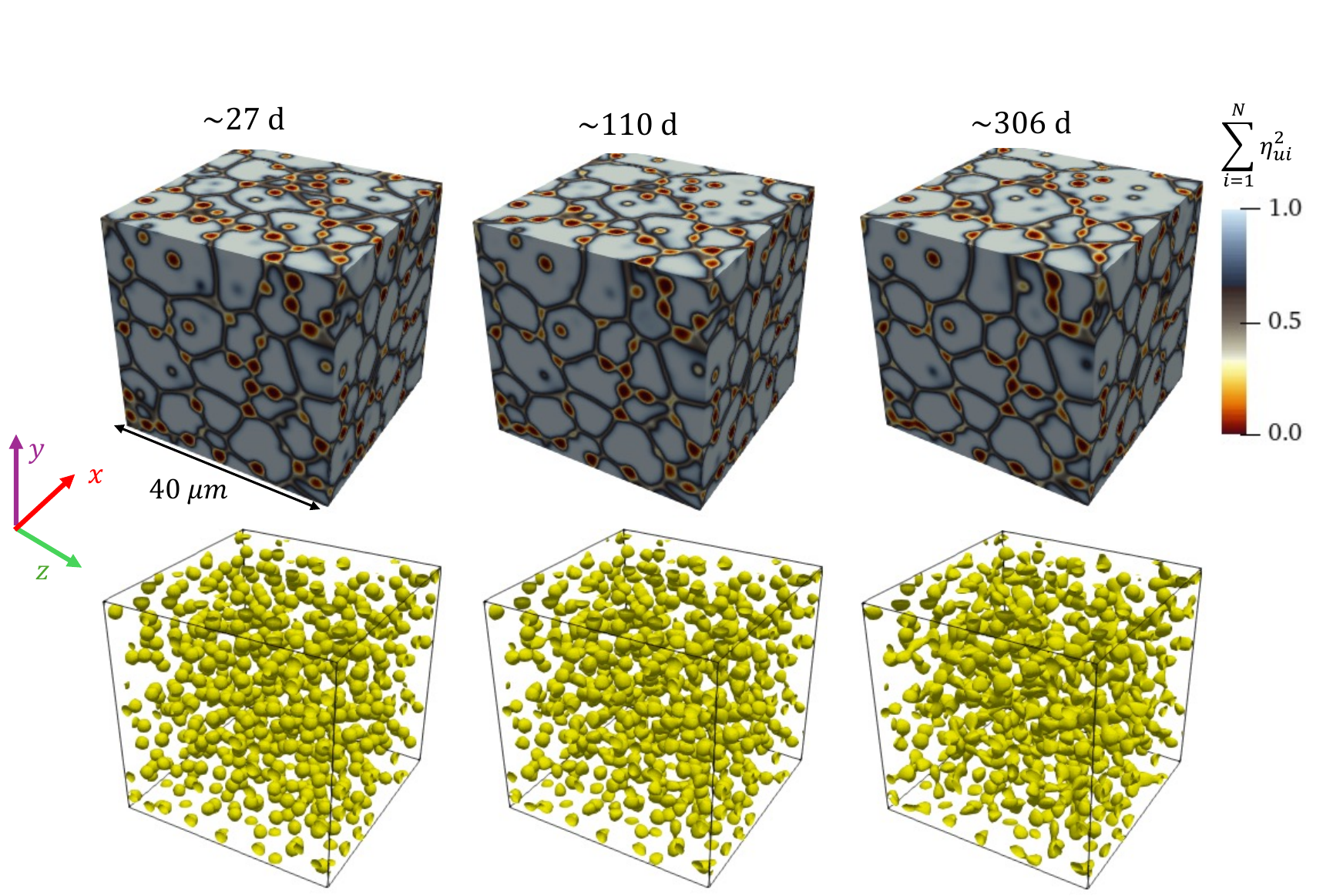}
     \caption{} \label{fig:evol_100gr_1600K}
 \end{subfigure}
 \caption{Evolution with time of 100-grain UO$_2$ polycrystals without a FS with $600$ initial bubbles at (a)  1200 K and (b) 1600 K. The top images are shaded by the function  $\Psi=\sum_{i=1}^{N}\eta_{ui}^2$, with light gray ($\Psi \approx 1$) representing grains, dark gray ($\Psi \approx 0.5$) representing GBs and red-yellow ($\Psi<0.4$) representing bubbles. The bottom images show intergranular bubble contour $(\eta_{b0}=0.5)$ in yellow. For clarity, the bottom images are shown without the initial GB planes and TJ lines.}
\label{fig:evol_100gr}
\end{figure}
\begin{figure}[tbp]
\centering
\begin{subfigure}{0.49\textwidth}
	\includegraphics[trim=0 0 0 0, clip, keepaspectratio,width=\linewidth]{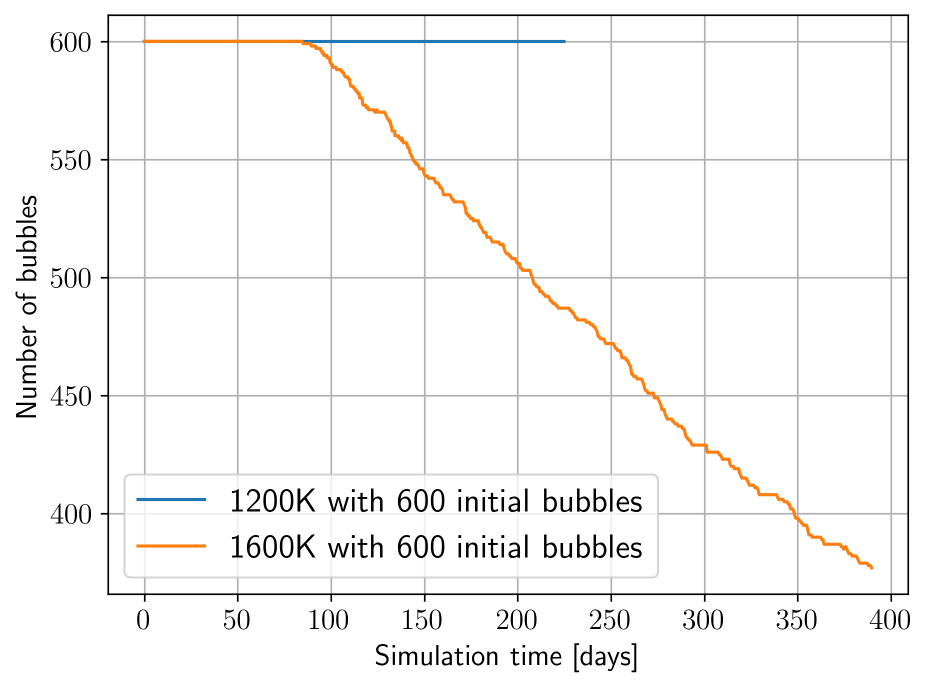}
	\subcaption{}
	\label{Fig8a}
\end{subfigure}
\begin{subfigure}{0.49\textwidth}
	\includegraphics[trim=0 0 0 0, clip, keepaspectratio,width=\linewidth]{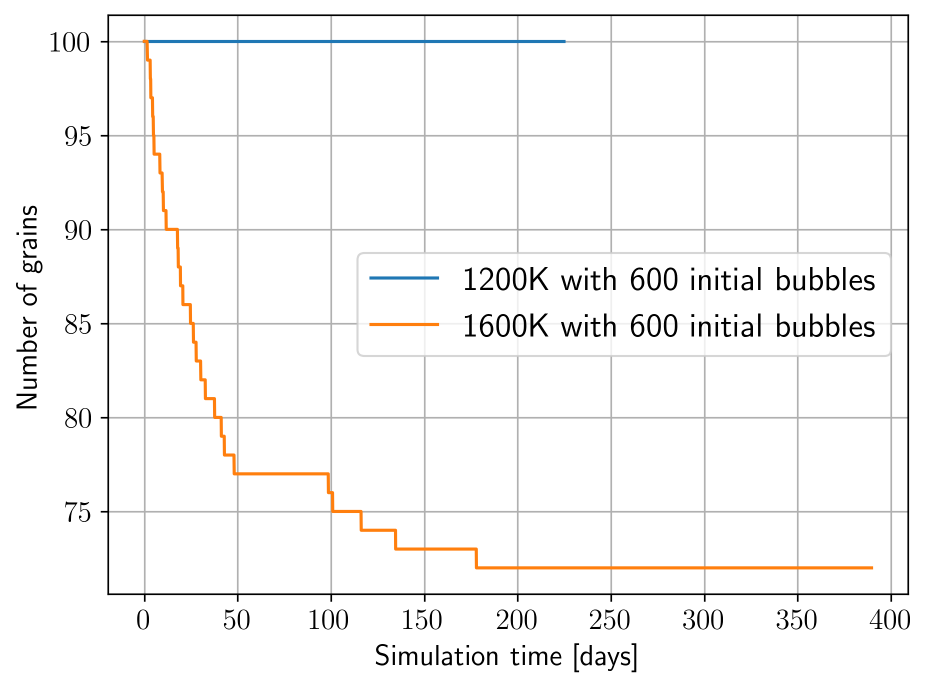}
	\subcaption{}
	\label{Fig8b}
\end{subfigure}
\begin{subfigure}{0.49\textwidth}
	\includegraphics[trim=0 0 0 0, clip, keepaspectratio,width=\linewidth]{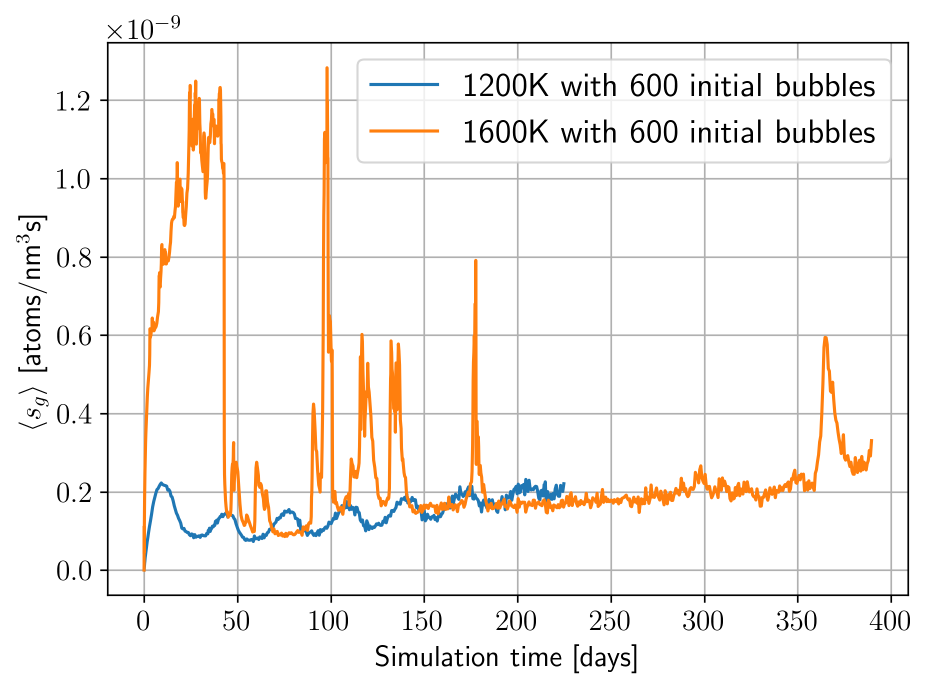}
	\subcaption{}
	\label{Fig8c}
\end{subfigure}
\begin{subfigure}{0.49\textwidth}
	\includegraphics[trim=0 0 0 0, clip, keepaspectratio,width=\linewidth]{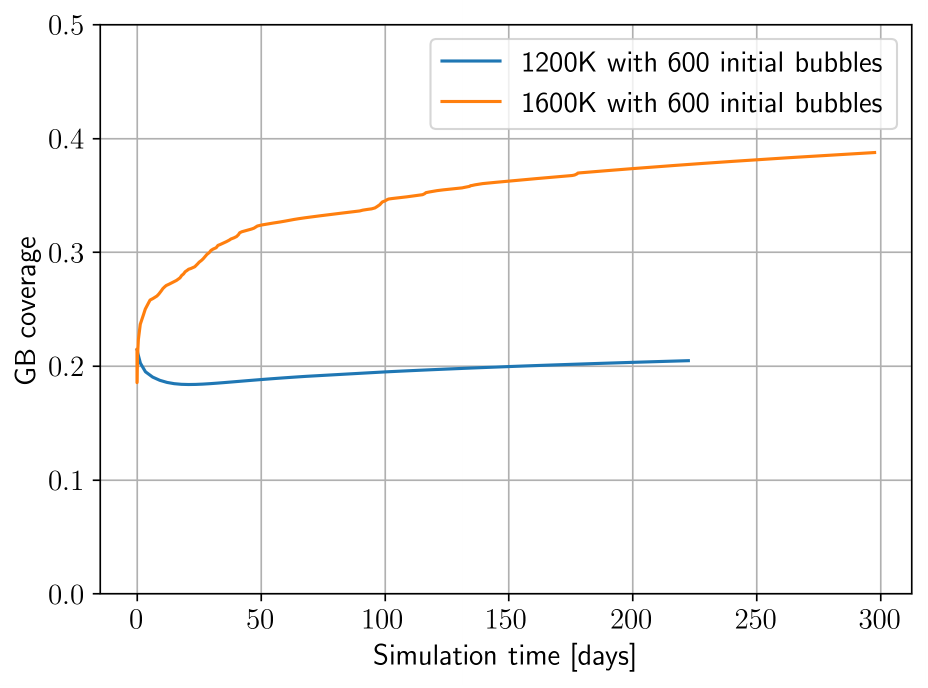}
	\subcaption{}
	\label{Fig8d}
\end{subfigure}
\centering
\begin{subfigure}{0.49\textwidth}
	\includegraphics[trim=0 0 0 0, clip, keepaspectratio,width=\linewidth]{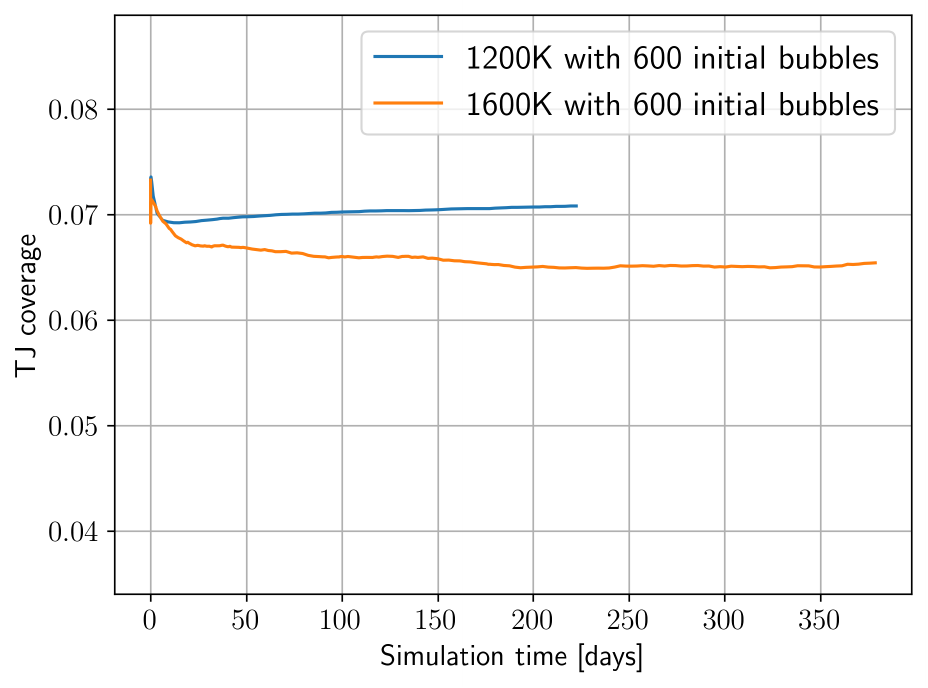}
	\subcaption{}
	\label{Fig8e}
\end{subfigure}
\caption{Metrics evolution with time quantifying the microstructure evolution in the $100$-grain polycrystals without a FS, where (a) shows the number of intergranular bubbles, (b) number of grains, (c) $\langle s_{g}\rangle$, or the volume-averaged Xe reaching GBs and intergranular bubbles surfaces, calculated using Eq.~\eqref{eq:av_sg}, (d) GB fractional coverage and (e) TJ fractional coverage.}
\label{Figure8}
\end{figure}

%
While the 10-grain simulations provide valuable insights into the interaction between intergranular bubbles and grains, the number of grains is still relatively small. We now simulate the GB and intergranular bubble evolution in 100-grain polycrystals to provide a broader view of the evolution that also focuses on Stage 3 of FGR behavior, starting from the initial condition shown in Fig.~\ref{Fig1}c.  Images of the evolving microstructures are shown in Fig.~\ref{fig:evol_100gr}. We visualize the GBs and bubbles on the outer boundaries of the 3D domain and we show just the intergranular bubbles. The metrics quantifying the microstructure evolution are shown in Fig.~\ref{Figure8}. 

Figure \ref{fig:evol_100gr_1200K} shows the bubble and GB evolution in the 100-grain simulation at $1200$ K. As in the $10$-grain simulations, both intergranular bubbles and GBs hardly evolve at $1200$ K. Both the number of bubbles and grains remain unchanged at $1200$ K after $\sim220$ days (Figs.~\ref{Fig8a} and \ref{Fig8b}). Note that there was a slight reduction in the number of bubbles in the 10-grain polycrystal at 1200 K, due to coalescence as the bubbles evolve towards a lenticular shape. However, since the initial bubble spacing (see Section \ref{setup_details}) in the 100-grain simulation is nearly twice that of the 10-grain simulation, no coalescence is observed at 1200 K. As noted previously, this is due to the low gas diffusivity and GB mobility at $1200$ K. 

In contrast, we observe simultaneous grain growth and bubble coalescence at $1600$ K  (Fig.~\ref{fig:evol_100gr_1600K}). The bubbles quickly take on a lenticular shape and GBs not pinned by bubbles quickly migrate. The number of bubbles does not change for the first 80 days, but then decreases by $\sim33.67$\%, from $600$ to $398$, in $350$ days, as shown in Fig.~\ref{Fig8a}. This is different than the behavior of the 10-grain polycrystals at 1600 K and is likely due to the larger initial radius (resulting in less bubble coarsening) and a larger spacing between bubbles (resulting in less coalescence). The number of grains drops very quickly at first, going from 100 to 77 in 50 days, but slows, dropping to 72 after 175 days and not changing further up to 350 days (see Fig.~\ref{Fig8b}). The GBs not pinned by bubbles undergo rapid migration until they reach bubbles and become pinned. 

As noted in Section~\ref{sec:10-grain}, grain growth leads to sweeping of the immobile Xe clusters by the UO$_2$ GBs, which increases the arrival rate of the gas atoms at GBs and bubble surfaces. At 1200 K, the average source term at the GBs and bubble surfaces is small due to the slow diffusion, as shown in Fig.~\ref{Fig8c}, and has no significant spikes because no GB migration occurs. It is much higher at 1600 K, and spikes in the average source term coincide with decreases in the number of grains. The large amount of gas arriving at the GBs and intergranular bubbles from 0 to 50 days causes the bubbles to grow, resulting in their eventual coalescence, causing the number of bubbles to go down.

Figure \ref{Fig8d} shows the GB coverage as a function of time for the 100-grain simulations at 1200 and 1600 K. As expected, the GB coverage barely changes at $1200$ K. However, at $1600$ K, the GB coverage increases from $21\%$ to $38\%$ in $\sim300$ days. It is worth noting that, in contrast to the 10-grain simulation at $1600$ K, the GB coverage in the 100-grain case continues to increase even after $350$ days. This suggests that GB coverage has not yet reached a high enough value to saturate due to coalescence. This is likely because the initial bubble density in the 100-grain simulations is lower than that in the 10-grain simulations, which can be attributed to the fact that the initial bubble spacing in the 100-grain case is double that of the 10-grain case, as noted in Section \ref{setup_details}. 

The TJ coverage marginally evolves at $1200$ and $1600$ K, as shown in Fig.~\ref{Fig8e}. As noted previously, this is because there are few initial TJ bubbles in our simulations, as evident by the relatively low ($\sim7$\%) initial TJ coverage. The TJ coverage marginally decreases in the 100-grain simulations at $1600$ K. However, this result may be an inaccurate representation of the actual TJ coverage because the TJ coverage calculation decreases in accuracy with time due to our use of the initial TJ lines in the calculation of the TJ coverage.


\subsection{Polycrystals with a FS}
\label{sec:results_FS}

\begin{figure}[tbp]
 \centering
 \begin{subfigure}{0.9\linewidth}
    \includegraphics[width=\linewidth]{Figure3.pdf}
    \caption{}
 \end{subfigure}
 \begin{subfigure}{0.9\linewidth}
    \includegraphics[width=\linewidth]{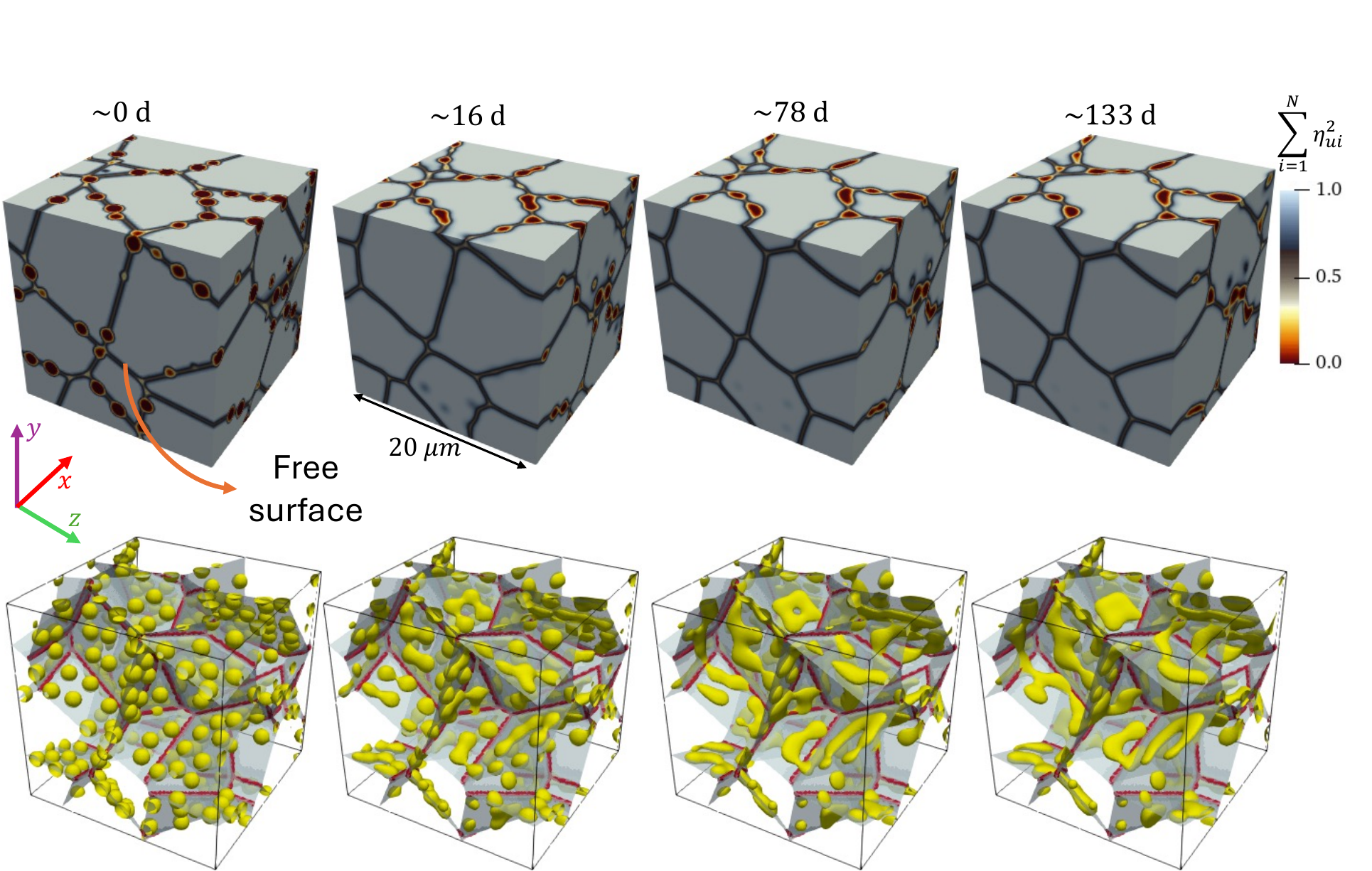}
    \caption{}\label{fig:evol_10gr_FS_sub}
 \end{subfigure}
 \caption{Evolution with time of 10-grain UO$_2$ polycrystals with 320 initial bubbles at 1600 K, where (a)  is without a FS (repeated from Fig.~\ref{fig:evol_10gr_320_1600K}) and (b) is with a FS. The top images are shaded by the function  $\Psi=\sum_{i=1}^{N}\eta_{ui}^2$, with light gray ($\Psi \approx 1$) representing grains, dark gray ($\Psi \approx 0.5$) representing GBs and red-yellow ($\Psi<0.4$) representing bubbles. In the bottom images the intergranular bubble contour ($\eta_{b0}=0.5$) is yellow, the initial GBs without bubbles are light gray, and the initial TJ lines without bubbles are red.}
\label{fig:evol_10gr_FS}
\end{figure}

\begin{figure}[tbp]
 \centering
 \begin{subfigure}{0.9\linewidth}
    \includegraphics[width=\linewidth]{Figure7.pdf}
    \caption{}
 \end{subfigure}
 \begin{subfigure}{0.9\linewidth}
    \includegraphics[width=\linewidth]{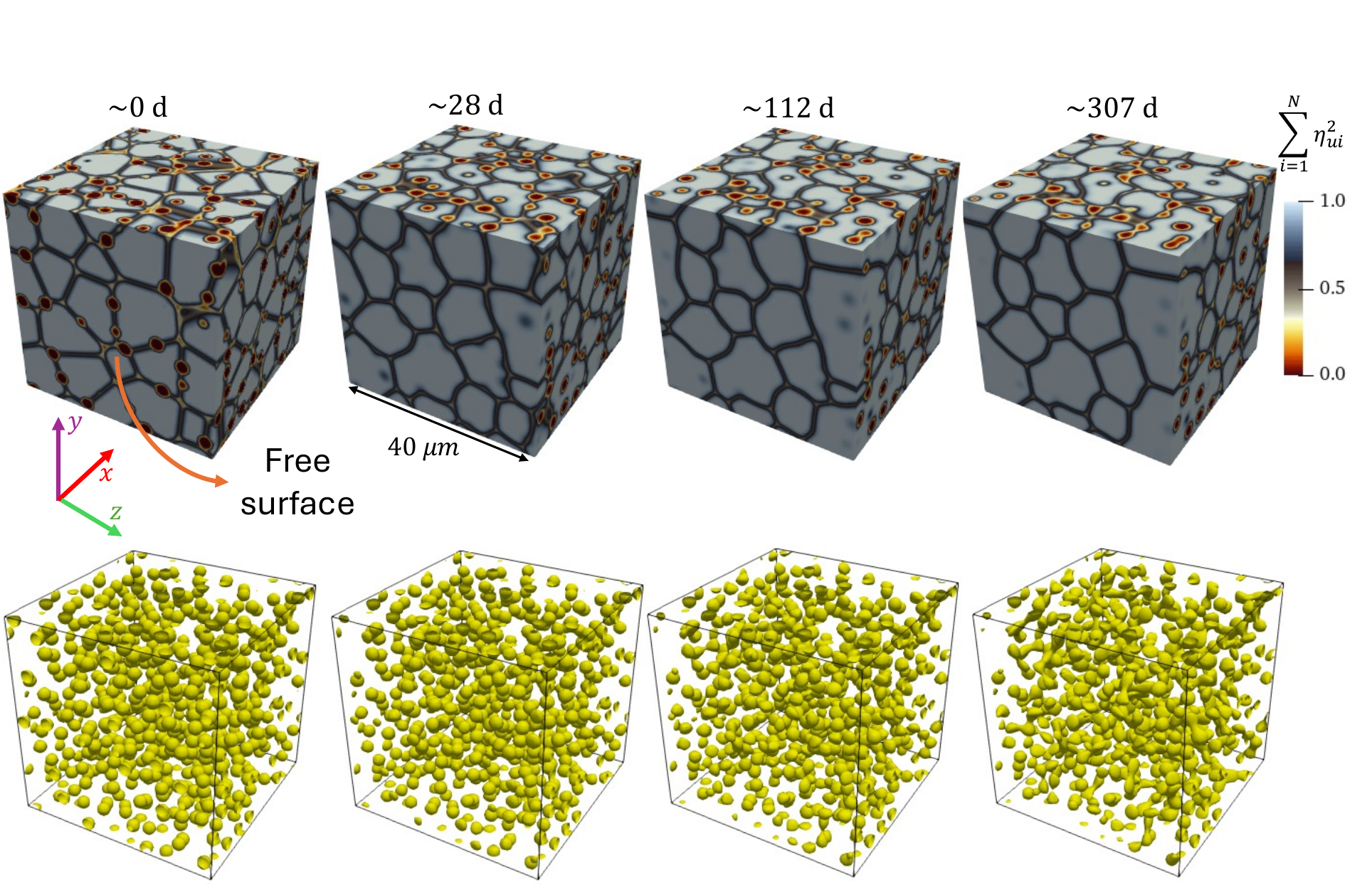}
    \caption{}\label{fig:evol_100gr_FS_sub}
 \end{subfigure}
 \caption{Evolution with time of 100-grain UO$_2$ polycrystals with 600 initial bubbles at 1600 K, where (a) is without a FS (repeated from Fig.~\ref{fig:evol_100gr_1600K}) and (b) is with a FS. Note that the simulations with and without a FS are performed using the re-solution model of Setyawan~\cite{setyawan2018atomistic} and Turnbull~\cite{pastore2023}, respectively. The top images are shaded by the function  $\Psi=\sum_{i=1}^{N}\eta_{ui}^2$, with light gray ($\Psi \approx 1$) representing grains, dark gray ($\Psi \approx 0.5$) representing GBs and red-yellow ($\Psi<0.4$) representing bubbles. The bottom images show intergranular bubble contour $(\eta_{b0}=0.5)$ in yellow. For clarity, the bottom images are shown without the initial GB planes and TJ lines.}
\label{fig:evol_100gr_FS}
\end{figure}

\begin{figure}[tbp]
\centering
\begin{subfigure}{0.49\textwidth}
	\includegraphics[trim=0 0 0 0, clip, keepaspectratio,width=\linewidth]{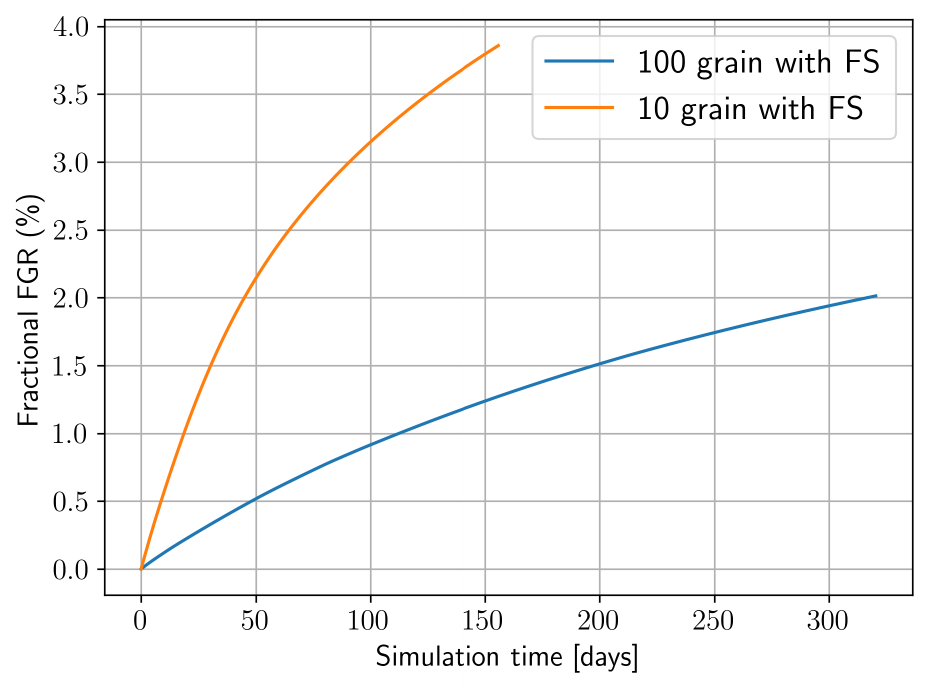}
	\subcaption{}
	\label{Fig11a}
\end{subfigure}
\begin{subfigure}{0.49\textwidth}
	\includegraphics[trim=0 0 0 0, clip, keepaspectratio,width=\linewidth]{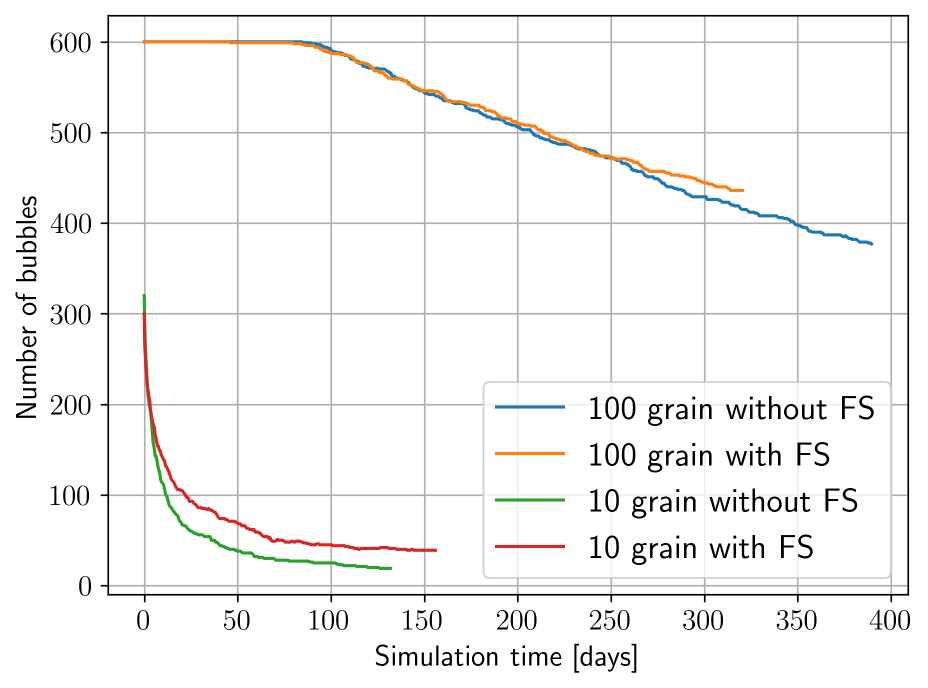}
	\subcaption{}
	\label{Fig11b}
\end{subfigure}
\begin{subfigure}{0.49\textwidth}
	\includegraphics[trim=0 0 0 0, clip, keepaspectratio,width=\linewidth]{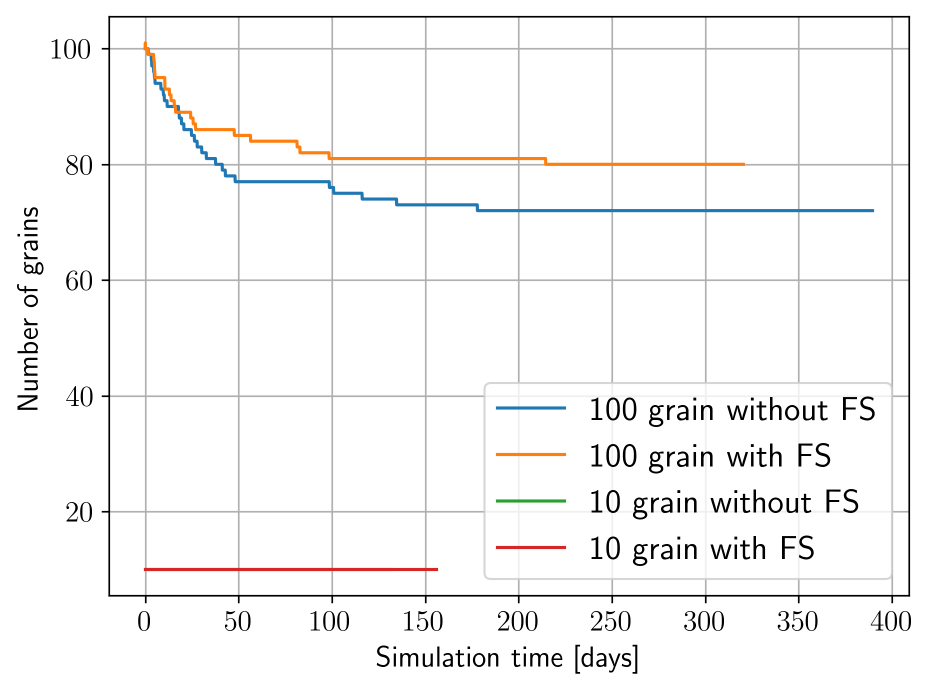}
	\subcaption{}
	\label{Fig11c}
\end{subfigure}
\begin{subfigure}{0.49\textwidth}
	\includegraphics[trim=0 0 0 0, clip, keepaspectratio,width=\linewidth]{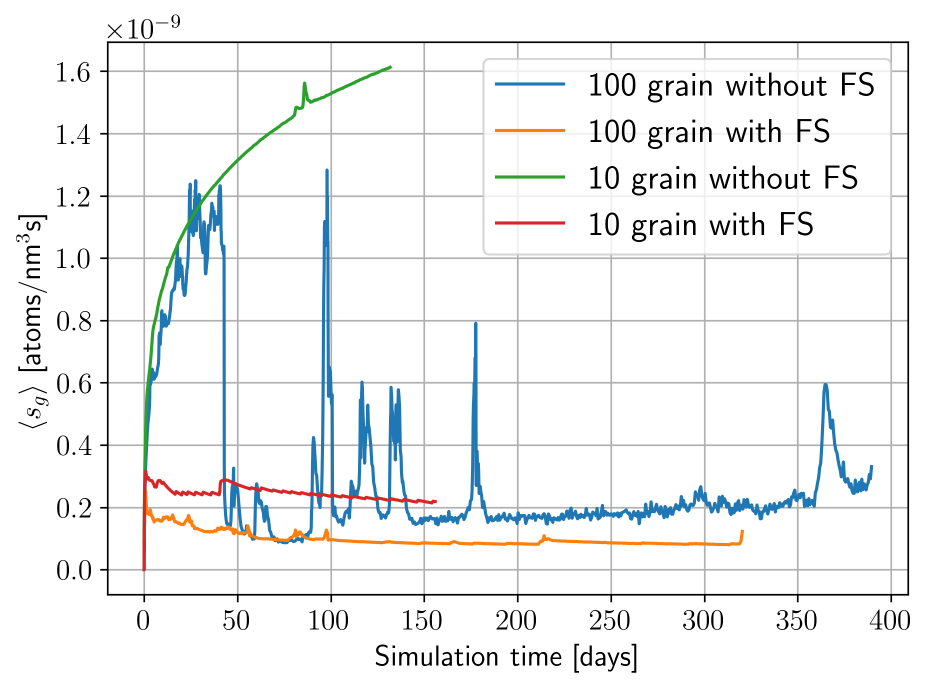}
	\subcaption{}
	\label{Fig11d}
\end{subfigure}
\begin{subfigure}{0.49\textwidth}
	\includegraphics[trim=0 0 0 0, clip, keepaspectratio,width=\linewidth]{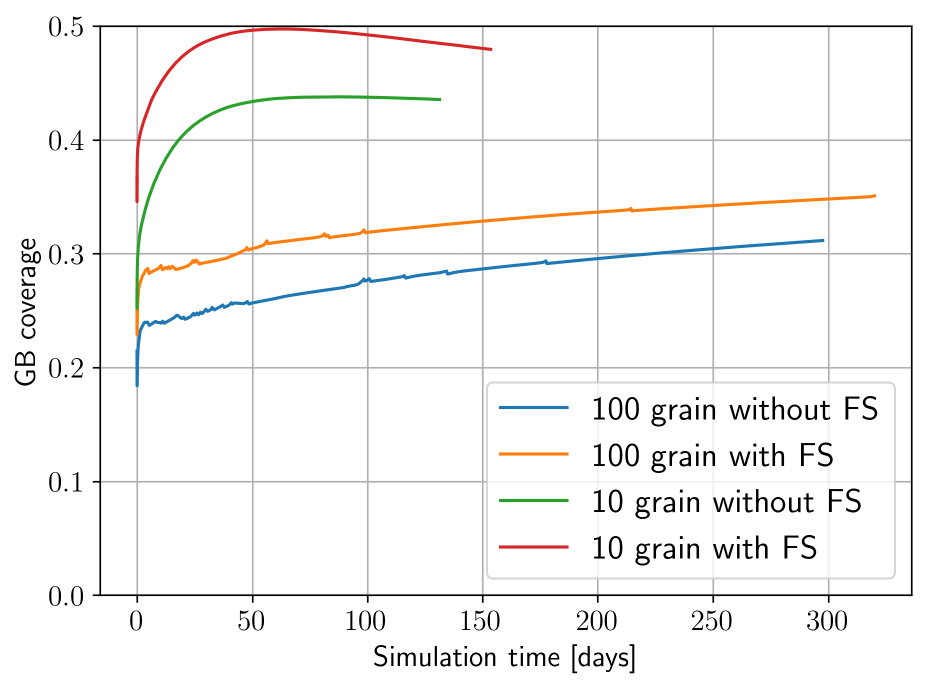}
	\subcaption{}
	\label{Fig11e}
\end{subfigure}
\begin{subfigure}{0.49\textwidth}
	\includegraphics[trim=0 0 0 0, clip, keepaspectratio,width=\linewidth]{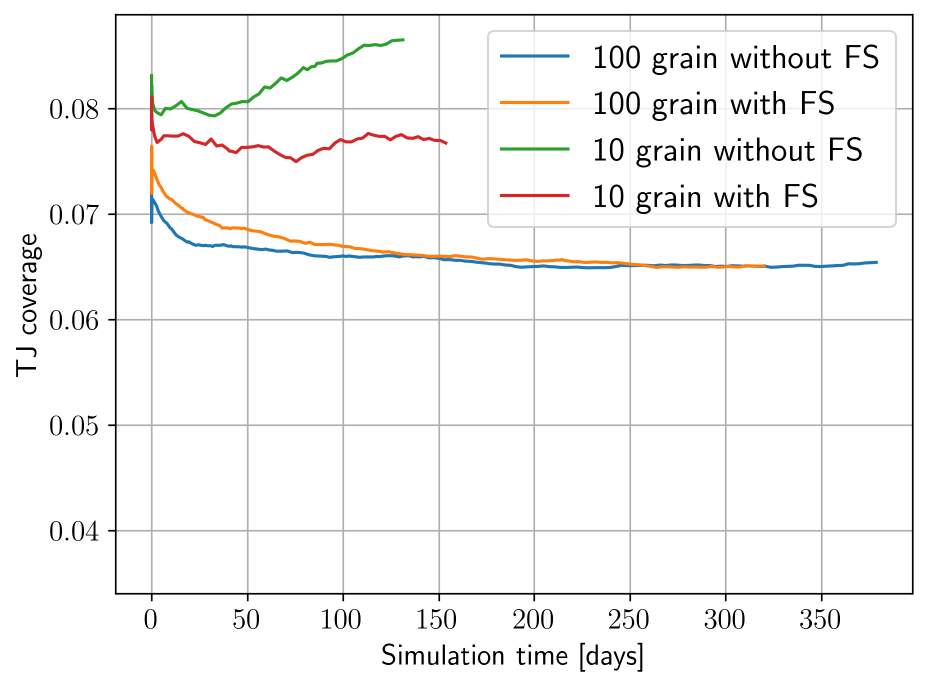}
	\subcaption{}
	\label{Fig11f}
\end{subfigure}
\caption{Metrics evolution with time quantifying the microstructure evolution in 10- and $100$-grain polycrystals with and without a FS, where (a) shows the fractional gas release, (b) number of intergranular bubbles, (c) number of grains, (d) $\langle s_{g}\rangle$, or the volume-averaged Xe reaching GBs and intergranular bubbles surfaces, calculated using Eq.~\eqref{eq:av_sg}, (e) GB fractional coverage, and (f) TJ fractional coverage.}
\label{Figure11}
\end{figure}

Muntaha \textit{et al.}~\cite{muntaha2023} predicted fission gas release (FGR) in 2D simulations by including a FS in their simulations. However, their simulations could not consider gas release through interconnected GB bubbles, since the GBs in 2D are just lines. Thus, here we predict the amount of gas released at 1600 K in a 10-grain UO$_2$ polycrystal with 320 initial intergranular bubbles and in a 100-grain polycrystal with 600 initial bubbles. The evolving microstructures with a FS are compared with those without a FS in Fig.~\ref{fig:evol_10gr_FS} for the 10-grain polycrystal and in Fig.~\ref{fig:evol_100gr_FS} for the 100-grain polycrystal. The metrics quantifying the evolution are shown in Fig.~\ref{Figure11}.

In the simulations with a FS, both intragranular gas in Xolotl and intergranular gas in MARMOT are released. Figure \ref{Fig11a} shows that the fractional FGR increases rapidly at first, but slows with time as the gas from the initial bubbles escapes. Also, the FGR is significantly higher in the 10-grain simulation than in the 100-grain simulation. This is because a larger fraction of the bubbles in the domain are in contact with or near the FS in the smaller 10-grain polycrystal than in the larger 100-grain polycrystal. Thus, much more of the gas can quickly diffuse through the bubbles and GBs to the FS in the 10-grain case. The gas escaping from the FS has a large impact on the bubble and GB evolution near the FS. In the 10-grain polycrystal (Fig.~\ref{fig:evol_10gr_FS_sub}), the bubbles at the FS quickly disappear, unlike in the case without a FS. The removal of those bubbles allows some GB migration to take place. Similar behavior occurs in the 100-grain polycrystal (Fig.~\ref{fig:evol_100gr_FS_sub}).

Figure~\ref{Fig11b} shows that the bubble loss rate is slightly slower in the 10-grain polycrystal with a FS than without it, although the initial number of bubbles is the same. This seems somewhat surprising, since the bubbles at the FS disappear due to FGR. However, since less gas is in the system, the bubbles grow and coalesce at a slower rate. The higher amount of coalescence without a FS causes the number of bubbles to drop faster than the bubbles disappearing due to FGR with a FS. The reduction in the number of bubbles is also slower in the 100-grain simulation, but not until the simulation time reaches ~250 days. This is likely due to the slower bubble coalescence in the 100-grain simulations than in the 10-grain simulations.

While Fig.~\ref{fig:evol_10gr_FS_sub} shows a small amount of GB evolution due to the disappearance of the bubbles at the FS, it is not enough to result in a drop in the number of grains, as shown in Fig.~\ref{Fig11c}. Thus, the number of grains is constant in the 10-grain cases with and without a FS. Note that the initial grain structures with and without a FS are somewhat different due to the impact of the boundary conditions on the Voronoi tesselation used to generate the grains. In the 100-grain polycrystal, the drop in the number of grains is smaller with a FS than without it. This is likely due to differences in the initial grain structure. 

Figure~\ref{Fig11d} shows that the gas arrival rate at GBs is much lower in 10- and 100-grain simulations with a FS than in the simulations without. Spikes in the 100-grain curve with a FS seem to correlated with GB migration, similar to the behavior without a FS, but the spikes are much smaller in magnitude. This is attributed to the difference in the re-solution models employed in the cases with a FS than without it, as noted in Section~\ref{CD_model}. 

The initial GB coverage values are higher in the simulations with a FS than without, as shown in Fig.~\ref{Fig11e}. The number of initial intergranular bubbles is the same, so this is due to the difference in the grain structures that results from the impact of the different boundary conditions on the Voronoi tesselation used to create the initial structures. The initial GB area is marginally higher in the domains with all periodic BCs than in the domains that are not periodic in the $x$-direction due to the FS. In the 10-grain polycrystal with and without a FS, the GB coverage increases rapidly at the start of the simulation but then substantially slows. Without a FS, the coverage saturates at around 43\%; with a FS, the coverage reaches nearly 50\%, and then it decreases. This decrease is likely due to the loss of gas atoms from the domain via FGR. In the 100-grain polycrystal, the GB coverages increase by roughly the same rate in the simulations with and without a FS. However, they do not get high enough to saturate.

The TJ coverage in the simulations with a FS start at similar values to those without a FS and does not change significantly, also like the case without a FS, as shown in Fig.~\ref{Fig11f}. In the 10-grain case, the TJ coverage increases slightly with a FS while it stays fairly constant without a FS. The TJ coverage in the 100-grain polycrystals is similar both with and without a FS.

\section{Discussion and future directions} 
\label{sec:discussion}
As noted in Section~\ref{PF_Model}, We have applied our hybrid fission gas model to investigate intergranular bubble evolution in large 3D polycrystals. This model captures many of the important mechanisms governing fission gas bubble behavior, but we also have to make some simplifying assumptions. First, we do not model nucleation of intergranular bubbles since the focus of our work is on understanding bubble evolution not nucleation; however, the initial bubble sizes and locations did impact our results, especially the TJ bubble behavior. Future studies should incorporate discrete nucleation in the PF model, including having intragranular bubbles in Xolotl act as nuclei in the PF model. Second, fission gas atoms in the PF model are assumed to always reside in a vacancy, so that when they escape from a FS a vacancy also escapes, causing the pore to shrink. In real fuel, gas can escape a bubble network without the pore network shrinking. This should be changed in future work to allow us to investigate the behavior of the underlying pore network once the gas escapes. Also, the current hybrid model does not consider the impact of gas pressure on the bubble network. This pressure is likely small in the larger intergranular bubbles but could be significant in the intragranular bubbles represented by Xolotl. Finally, we used fairly large interfacial widths in the PF model to reduce the computational expense. Future simulations should use smaller widths, though it would make the computational cost even higher than it already is for these simulations. 

Despite these limitations, our simulations represent a significant enhancement over past PF simulations of fission gas bubble evolution. First, they are the first application of the hybrid model to simulate fission gas in 3D for more than just a couple of hours of simulation time. In fact, to our knowledge, they are the largest mesoscale simulations of intergranular FG bubble behavior carried out to date. The hybrid model allows us to accurately capture the impact of intragranular bubble trapping on the bubble behavior and to include important mechanisms such as GB sweeping, as illustrated by the arrival rate of gas atoms at interfaces shown in Figs~\ref{Fig4c}, \ref{Fig8c}, and \ref{Fig11c}. In addition, we have carried out the first 3D PF simulations of intergranular bubbles interacting with evolving 3D GB structures; past simulations used either bicrystals \cite{Millet_2012_a,lan2024three}, hexagonal grains \cite{Larry_2019}, columnar grains \cite{lan2024three}, or static grain structures that do not evolve \cite{PRUDIL2022153777}. Due to our use of 3D grain structures, we were able to see the interactions between intergranular bubble evolution, GB migration and pinning, and FGR. For example, a smaller number of intergranular bubbles resulted in more GB migration, as shown in Fig.~\ref{Figure4}. Finally, we used boundary conditions to make one side of the domain into a FS for FGR, allowing us to see interactions such as bubble collapse from FGR resulting in additional GB migration, as shown in Fig.~\ref{fig:evol_10gr_FS}.

In this work, we started by validating our 3D hybrid simulations by comparing the evolution of the bubble density with the mean bubble area with that from experimental data taken from White \cite{White_2004}, as shown in Fig.~\ref{fig:validation}. Similar comparisons have been done in the past using the PF method to model FG bubble evolution \cite{Millet_2012_a,PRUDIL2022153777}. White's analytical model from Eq.~\eqref{Eq:White_Analytical} relates the change in bubble density and mean bubble area due to coalescence and is slightly to the right of his data, as we mention in Section \ref{sec:results_validation}. The past PF simulation results \cite{Millet_2012_a,PRUDIL2022153777} all start to the left of the curve, since their initial bubbles have to grow due to fission gas generation before they come into contact and begin to coalesce (see Fig.~\ref{fig:validation}). However, they eventually end to the right of the curve. All of our simulations at 1600 K also begin to the left of the curve and stays left of the curve while the bubbles grow. Our 10-grain simulation with 160 initial bubbles and 100-grain simulation with 600 initial bubbles always stay far to the left of the curve because the bubbles never experience significant coalescence. The 10-grain 320 initial bubble simulation at 1600 K experiences a large amount of bubble coalescence and does approach the coalescence curve, though it always stays to the left. The cause for why past PF simulations end to the right of the curve while ours stay to the left, similar to the experimental data, is unclear. Our simulations differ from the previous PF models \cite{Millet_2012_a,PRUDIL2022153777} primarily because we include GB migration in our model. Therefore, one possible reason is that GB evolution helps to speed bubble coalescence, decreasing the bubble area for a given bubble density and causing them to be to the left of the coalescence curve.

We performed simulations for both the 10- and 100-grain polycrystals at 1200 K, representing a mid-radius region of the fuel pellet, and 1600 K, representing a region near the center of the pellet. All our simulations show very little evolution at 1200 K; only experiencing slight GB migration and bubble shape change (see Figs.~\ref{fig:evol_10gr_320_1200K} and \ref{fig:evol_100gr_1200K}). The simulations at 1600 K show large amounts of evolution, including bubble growth and coalescence, and GB migration when they are not pinned by bubbles (see Figs.~\ref{fig:evol_10gr_320_1600K}, \ref{fig:10gr_1600K_160b}, and \ref{fig:evol_100gr_1600K}, as well as metrics in Figs.~\ref{Figure4} and \ref{Figure8}). Our simulations with a FS at 1600 K show significant FGR (Fig.~\ref{Fig11a}). This is all consistent with the experimental finding that FGR initiates around 1500 K \cite{bagger1994temperature}.

We also performed 10-grain simulations with different initial bubble densities (see the initial GB bubble areal densities from our simulations in Fig.~\ref{fig:validation}): the 100 grain simulations have the smallest initial bubble density, then the 10-grain simulations with 160 initial bubbles, and the 10-grain simulations with 320 initial bubbles have the largest bubble density among the three. However, since it depends on both the number density and on the mean bubble area, the 10-grain 160 bubble simulations have the smallest initial GB coverage (0.05), then the 100-grain simualtions (0.19), and then the 10-grain 320 bubble simulations (0.22). Though the fission gas production rate is the same in all the simulations, the increase in the GB coverage in all the simulations at 1600 K depends somewhat on the initial GB coverage: it changes by 0.17 in the 10-grain 160 bubble simulation, and by 0.2 in both the 100-grain and 10-grain 320 bubble simulations. This is because when the initial GB coverage is larger, the bubble surfaces are closer together, so they coalesce easier. The change in the grain growth varies with initial bubble density rather than grain boundary coverage. The 100-grain simulation experiences the most grain growth, then the 10-grain 160 bubble simulation, and the 10-grain 320 simulation experiences no grain growth. This is because more smaller bubbles on a GB have a larger pinning force than one large bubble, even if they have the same GB coverage. This is consistent with the results from the analytical grain growth model presented by Tonks \emph{et al.}~\cite{tonks2021mechanistic} that couples the grain size with the pore number density rather than the GB coverage.

It is commonly accepted that the formation of TJ tunnels is an important part of FGR. TJ tunnel networks are assumed to form earlier than GB bubble networks, and so once a GB network connects to a TJ network, the gas can quickly move through the fuel in these networks. Our simulations had only small TJ coverages and it did not significantly increase. However, these results should not be taken as evidence that TJ bubbles are not important in FGR because the accuracy of our TJ coverage calculation decreases with time in simulations that experience significant GB migration and because we did not place initial bubbles at TJs. Our method to calculate the TJ coverage uses the initial TJ locations for calculating the total TJ length, as described in Appendix~\ref{AppendixA2}. In the 1200 K simulations and the 1600 K 10-grain 320 bubble simulation, this should not impact the results because the GBs do not migrate. The 1600 K 10-grain 160 bubble and 100-grain simulations do experience significant GB migration, and so the total TJ length we use will get less and less accurate with time. More significantly, we randomly introduce our initial bubbles across the GB surfaces, with no preference for TJs. However, it is likely that TJs will have more free volume than many GBs and so vacancies and gas atoms will preferentially segregate to them over GBs. This would result in earlier nucleation of bubbles at TJs. Future simulations should be carried out that include the nucleation of bubbles and preferential segregation to TJs to more accurately represent the impact of TJs on FG behavior. 
In addition, experiments are needed to characterize TJ bubbles; So far, no experimental data have been reported on TJ bubble density and coverage in irradiated polycrystalline UO$_2$~\cite{Larry_2019}. 

We performed two simulations that include a FS, allowing us to compute the FGR, as shown in Fig.~\ref{Fig11a}. Looking at the fuel microstructures, the FS causes the initial bubbles at and near the FS to shrink and eventually disappear (see Fig.~\ref{fig:evol_10gr_FS_sub} and \ref{fig:evol_100gr_FS_sub}). Thus, compared to the simulations without a FS, we see fewer interconnected bubble networks, and we never see bubble networks near the FS. Our simulations form a denuded zone free of GB and TJ bubbles near the FS that could prevent interconnected bubble networks that connect to the FS from forming. In fact, at an initial FS of a fuel pellet, such as the pellet top or bottom, intergranular bubbles would never form because the FG would simply escape. The case simulated here, where initial bubbles exist at a FS, is similar to the formation of a new crack surface that is in contact with isolated intergranular bubbles. It would be interesting in the future to simulate with a FS where the initial condition is the end state of a simulation of bubble evolution at 1600 K without a FS, like those we show in Figs.~\ref{fig:10gr_1600K} and \ref{fig:evol_100gr_1600K}. Then, we could simulate the behavior when an interconnected network of bubbles comes in contact with a FS, rather than just isolated bubbles like we have in these simulations.

\section{Conclusions} \label{sec:conclusions}
In this study, we used a previously developed hybrid multiscale approach~\cite{DongUk-Kim,muntaha2023} to understand the influence of fuel temperature, initial bubble density, and grain structure on intergranular bubble evolution and FGR in UO$_2$. The novelty of this approach is that it models the fast gas diffusion along UO$_2$ GBs and bubble surfaces using the PF approach, as well as the trapping and re-solution of gas atoms within UO$_2$ grains using the cluster dynamics approach. In contrast to the previous 3D study \cite{muntaha2023}, which was limited to five grains and a couple of hours of simulation time, our simulations predict both intergranular bubble evolution and FGR in 10-grain and 100-grain microstructures for months. In addition, unlike Muntaha \emph{et al.}~\cite{muntaha2023}, we calculate GB coverage with time, a widely used criterion for FGR in engineering-scale codes.

Temperature sharply partitions behavior: at 1200 K, evolution is minimal, while at 1600 K bubbles become lenticular and coalesce as unpinned GBs migrate until pinned. Bubble density versus mean projected area follows White’s experimental coalescence behavior and remains to the left of the analytical curve, plausibly because GB migration accelerates coalescence. Initial bubble density governs the balance between pinning and grain growth; fewer bubbles reduce coalescence but permit more GB migration and transient spikes in interfacial gas arrival due to GB sweeping. With a FS, early release is rapid and bubbles at or near the surface collapse to form a denuded zone, suppressing local network connectivity and slowing subsequent coalescence; GB coverage increases at 1600 K and saturates below 50\%. TJ coverage remains low without preferential nucleation at TJs, so TJ‑mediated transport is underestimated unless TJ‑biased nucleation and improved TJ tracking during GB migration are included.

Future work should incorporate discrete nucleation with possible TJ preference, decouple gas escape from pore shrinkage, include gas pressure effects (particularly for intragranular bubbles), reduce phase‑field interface widths to sharpen quantitative predictions, and simulate free‑surface contact with pre‑formed interconnected networks. Overall, these simulations, the largest 3D mesoscale studies of intergranular fission gas behavior to date, provide mechanistic insight and quantitative metrics that can inform and refine engineering‑scale FGR models.

%% file: Appendix_A.tex
\section*{Appendix A}
\label{Appendix_A}

\setcounter{equation}{0} 
\renewcommand{\theequation}{A.\arabic{equation}}

\setcounter{subsection}{0} 
\renewcommand{\thesubsection}{A.\arabic{subsection}}

\subsection{Expressions for grand-potential densities}
\label{AppendixA1}
In this section, we provide explicit expressions for the grand-potential densities $\omega_{\theta=m,b}$ of the UO$_2$ matrix and bubble phases. To this end, we initially derive the phase atomic fractions, $c_{g}^{\theta=m,b}$ and $c_{v}^{\theta=m,b}$, as functions of the gas and vacancy chemical potentials from the free energy densities, $f_{\theta=m,b}$. Following Refs.~\cite{Larry_2018,DongUk-Kim, muntaha2023}, we assume the following parabolic expressions for the free energy densities:
\begin{align}
f_{\theta}\left(c_{g}^{\theta}, c_{v}^{\theta}\right) = \mathcal{A}_{g}^{\theta}/2 (c_{g}^{\theta} - c_{g}^{\theta, eq})^2 + \mathcal{A}_{v}^{\theta}/2 (c_{v}^{\theta} - c_{v}^{\theta,eq})^2, \quad \theta = \lbrace m, b\rbrace
\label{EqnA1.1}
\end{align}
where $\mathcal{A}_{g}^{\theta}$ and $\mathcal{A}_{v}^{\theta}$ are the parabolic coefficients, $c_{g}^{\theta, eq}$ and $c_{v}^{\theta,eq}$ are the equilbrium atomic fractions of Xe and Va within the bulk phases. By differentiating Eq.~\eqref{EqnA1.1} with respect to the phase atomic fractions $c_{k=g,v}^{\theta}$ and using $\mu_{k}=v_{a}\left(\partial f_{\theta}/\partial c_{k}^{\theta}\right)$, the phase atomic fractions, $c_{g}^{\theta=m,b}$ and $c_{v}^{\theta=m,b}$ can be explicitly expressed as~\cite{Larry_2018, Plapp_2011}:
\begin{align}
c_{g}^{\theta}\left(\mu_{g}\right) &= \frac{\mu_{g}}{v_{a}\mathcal{A}_{g}^{\theta}}+ c_{g}^{\theta,eq}, \quad \theta = \lbrace m, b\rbrace \label{EqnA1.2}\\
c_{v}^{\theta}\left(\mu_{v}\right) &= \frac{\mu_{v}}{v_{a}\mathcal{A}_{v}^{\theta}}+ c_{v}^{\theta, eq} \quad \theta = \lbrace m, b\rbrace.  \label{EqnA1.3}
\end{align} 
Substituting Eqs.~\eqref{EqnA1.2} and \eqref{EqnA1.3} in Eq.~\eqref{EqnA1.1} and using Eq.~\eqref{Eqn5} yields the following grand-potential densities of the matrix and bubble phase as functions of $\mu_{g}$ and $\mu_{v}$:
\begin{align}
\omega_{m}\left(\mu_{g},\mu_{v}\right) &= -\left(\frac{\mu_{g}^2}{2v_{a}^2\mathcal{A}_{g}^m} +  \frac{\mu_{g}}{v_{a}}c_{g}^{m,eq} +\frac{\mu_{v}^2}{2v_{a}^2\mathcal{A}_{v}^m} + \frac{\mu_{v}}{v_{a}}c_{v}^{m,eq}\right)\label{EqnA1.4}\\
\omega_{b}\left(\mu_{g},\mu_{v}\right) &= -\left(\frac{\mu_{g}^2}{2v_{a}^2\mathcal{A}_{g}^b} +  \frac{\mu_{g}}{v_{a}}c_{g}^{b,eq} +\frac{\mu_{v}^2}{2v_{a}^2\mathcal{A}_{v}^b} + \frac{\mu_{v}}{v_{a}}c_{v}^{b,eq}\right)\label{EqnA1.5}
\end{align}

Table~\ref{table1} provides the parabolic coefficients, $\mathcal{A}_{g}^{\theta}$ and $\mathcal{A}_{v}^{\theta}$, and the equilibrium atomic fractions, $c_{g}^{\theta, eq}$ and $c_{v}^{\theta,eq}$, required to formulate Eqs.~\eqref{EqnA1.2}, \eqref{EqnA1.3},\eqref{EqnA1.4}, and \eqref{EqnA1.5}. Similar to Refs.~\cite{Larry_2018,DongUk-Kim, muntaha2023}, we assume that $\mathcal{A}_{g}^{m} = \mathcal{A}_{v}^{m}$ and $\mathcal{A}_{g}^{b} = \mathcal{A}_{v}^{b}$. 

\subsection{Method to calculate GB and TJ coverages}
\label{AppendixA2}
The GB coverage is calculated by taking the ratio of the GB area occupied by bubbles to the total GB area without bubbles. In this study, the total GB area without bubbles is calculated analytically using $3V/2\langle R(t)\rangle$ \cite{Gottstein_Book}, where $V$ is the domain volume and $\langle R(t)\rangle$ is the mean grain radius, which is calculated from the mean grain volume assuming spherical grains. The mean grain volume is calculated using the MOOSE postprocessor \texttt{AverageGrainVolume}~\cite{PERMANN201618}. The GB area covered with bubbles is calculated from the area of the grain and bubble contours using ParaView software~\cite{ahrensparaview}.

TJ coverage is calculated by taking the ratio of the TJ volume occupied by bubbles to the total TJ volume without bubbles. For simplicity, we assume that the TJ volume without bubbles remains constant with time, which is a valid assumption for cases where no grain growth is observed. Thus, we use the initial microstructures to calculate the total TJ volume without bubbles as follows:

To quantify the initial TJ length in the simulation, we calculate the following function at time $t=0$:
\begin{align}
    \mathcal{R}(\boldsymbol{\eta}_{u}) = \sum_{i=1}^{N}\sum_{j=i+1}^{N}\sum_{k=j+1}^{N}\eta_{ui}^2\eta_{uj}^2\eta_{uk}^2.
\end{align}
If the function $\mathcal{R}(\boldsymbol{\eta}_{u}) $ exceeds $10^{-5}$, then the volume element belongs to a TJ, and the function is locally assigned a value of $1$.  By integrating this function over the domain volume, the total initial TJ length (or volume) is determined. Note that this approach requires the GBs to have a finite thickness at time $t=0$, which is assumed to be the same as specified in Table~\ref{table1}. To calculate the TJ volume occupied by bubbles, we track the regions where the bubble contour overlaps with the initial TJ lines.